\documentclass[10pt,journal,twocolumn]{IEEEtran}
\newif\ifCLASSOPTIONromanappendices \CLASSOPTIONromanappendicestrue
\usepackage{a0size}
\usepackage{amssymb}
\usepackage{multicol}
\usepackage[english]{babel}
\usepackage{epsfig}
\usepackage{bm}
\usepackage{amsfonts,color,amsthm,amsmath, lscape,graphics}
\usepackage{url}
\usepackage{algorithm}
\usepackage{algpseudocode}

\usepackage[font=footnotesize]{caption}
\usepackage[font=footnotesize]{subcaption}
\usepackage{subcaption}
\usepackage{epstopdf}
\usepackage{algorithm}
\usepackage{algpseudocode}
\usepackage{cite}

\usepackage{fix2col}

\newcommand{\bh}{\mathbf{h}}

\newcommand{\bI}{\mathbf{I}}

\newcommand{\bx}{\mathbf{x}}
\newcommand{\bX}{\mathbf{X}}

\newcommand{\bs}{\mathbf{s}}

\newcommand{\bA}{\mathbf{A}}

\newcommand{\bH}{\mathbf{H}}

\renewcommand{\frac}{\dfrac}

\newcommand{\X}{{\cal X}}

\addto\captionsenglish{}

\IEEEoverridecommandlockouts
\newcommand{\changet}[1]{{\color{black}#1}}
\newcommand{\changett}[1]{{\color{black}#1}}
\newcommand{\changem}[1]{{\color{black}#1}}
\newcommand{\changes}[1]{{\color{black}#1}}
\begin{document}
\title{One-Bit Precoding and Constellation Range Design for Massive MIMO with QAM Signaling}

\author{Foad~Sohrabi, \IEEEmembership{Student Member, IEEE}, Ya-Feng~Liu, \IEEEmembership{Member, IEEE} and~Wei~Yu, \IEEEmembership{Fellow,~IEEE}
\thanks{F. Sohrabi, and W. Yu are with The Edward S. Rogers Sr. Department
of Electrical and Computer Engineering, University of Toronto, Toronto,
ON M5S 3G4, Canada (e-mail: fsohrabi@comm.utoronto.ca; weiyu@comm.utoronto.ca).
Y.-F. Liu is with the State Key Laboratory of Scientific and Engineering
Computing, Institute of Computational Mathematics and Scientific/Engineering
Computing, Academy of Mathematics and Systems Science, Chinese Academy
of Sciences, Beijing 100190, China (e-mail: yafliu@lsec.cc.ac.cn). \changes{This paper has been presented in part at the Information Theory and Applications (ITA) Workshop, San Diego, CA, U.S.A. February 2018.}
This work is supported \changes{in part} by the Natural Science and Engineering Research Council
(NSERC) of Canada, a Steacie Fellowship, in part by Qualcomm Technologies,
Inc. and in part by the National Natural Science Foundation of China under Grants 11671419 and 11688101.}
}
\maketitle
%%%%%%%%%%%%%%%%%%%%%%%%%%%%%%%%%%%%
%% Abstract
%%%%%%%%%%%%%%%%%%%%%%%%%%%%%%%%%%%%
\begin{abstract}
The use of low-resolution digital-to-analog converters (DACs) for
transmit precoding provides crucial energy efficiency advantage for
massive multiple-input multiple-output (MIMO) implementation.
This paper formulates a quadrature amplitude modulation (QAM)
constellation range and one-bit symbol-level precoding design problem
for minimizing the average symbol error rate (SER) in 
downlink massive MIMO transmission.
A tight upper-bound for SER with low-resolution DAC precoding is first
derived.  The derived expression suggests that the performance
degradation of \changes{one-bit} precoding can be interpreted as a decrease in the
effective minimum distance of the QAM constellation.  Using the
obtained SER expression, we propose a QAM constellation range
design for the single-user case.  
\changem{It is shown that in the massive MIMO limit, a reasonable choice}
for constellation range with \changes{one-bit precoding} is that of the infinite-resolution precoding
with per-symbol power constraint, but reduced by a factor of \changett{$\sqrt{2/\pi}$} or about $0.8$. The corresponding
minimum distance reduction translates to about 2dB gap between
the performance of \changes{one-bit} precoding and infinite-resolution precoding. 
This paper further \changes{proposes a low-complexity} %two-step 
heuristic algorithm
for one-bit precoder design.
Finally, the proposed QAM constellation range and precoder design are generalized to the multi-user downlink. We propose to scale the constellation range for infinite-resolution zero-forcing (ZF) precoding with per-symbol power constraint
by the same factor of \changett{$\sqrt{2/\pi}$} for \changes{one-bit precoding}. The proposed \changes{one-bit} precoding scheme is shown to be within 2dB of \changes{infinite-resolution ZF}. In term of number of antennas, \changes{one-bit} precoding requires about 50\% more antennas to achieve the same performance as infinite-resolution precoding.
\end{abstract}

%%%%%%%%%%%%%%%%%%%%%%%%%%%%%%%%%%%%
%% I) Introduction
%%%%%%%%%%%%%%%%%%%%%%%%%%%%%%%%%%%%
\section{Introduction}
Massive multiple-input multiple-output (MIMO) systems in which the
base station (BS) is equipped with large-scale antenna arrays, e.g.,
in the order of several hundreds of antennas, is a promising candidate for the
next generation wireless systems for achieving high spectral
efficiency, reliability, and connectivity requirements
\cite{andrews2014will}.  One of the main challenges in designing the downlink
massive MIMO transmission scheme is that the implementation of the
conventional digital precoding strategies such as
\cite{shi2011iteratively,wiesel2008zero,Hayssam_Wei_2010,liu_2013} may not be practical, because 
conventional precoding schemes require one dedicated high-resolution
digital-to-analog converters (DACs) for each antenna element. When a
BS transmitter is equipped with large number of antennas, this leads
to high hardware complexity and excessive circuit power consumption. 

To address the above issue, one line of works
\cite{zhang2005variable,el2013spatially,sohrabi2016hybrid,liang2014low}
propose hybrid precoding architecture in which the precoder consists
of two parts: low-dimensional digital precoder and high-dimensional
analog precoder. In the hybrid precoding, the number of required DACs
scales with the number of data streams rather than the number of
antennas \cite{sohrabi2016hybrid}. Although the hybrid precoding
scheme has been shown to be capable of approaching the performance of
the conventional digital zero-forcing (ZF) precoding scheme
\cite{liang2014low}, hybrid precoding still requires high-resolution
DACs. As the power consumption of DAC grows exponentially with the
resolution \cite{walden1999analog,svensson2006power}, the use of
high-resolution DAC can lead to high power consumption, especially
when  
the BS needs to transmit a large number of data streams.

\changet{This paper considers an alternative precoding architecture called \changes{one-bit}
precoding in which two DACs (i.e., one DAC for converting the in-phase signal and the other one for converting the quadrature signal) are dedicated for each antenna element but
with only \changes{one-bit} resolution}.  The \changes{one-bit}
precoding enables us to reduce the circuit power consumption of the
DACs.
Moreover, the \changes{one-bit} precoding satisfies the condition
of constant envelop transmission, hence it can prevent the possible
amplitude distortions which may occur in the practical systems when
the power amplifiers work in the saturation region
\cite{larsson2013constantenv}.

%%%%%%%%%%%%%%%%%%%%%%%%%%%%%%%%%%%%
%% I-A) Main Contributions
%%%%%%%%%%%%%%%%%%%%%%%%%%%%%%%%%%%%
\subsection{Main Contributions}
This paper considers the precoder design problem for the downlink of
a multi-user multiple-input single-output (MISO) system where only
one-bit resolution DACs are available at the transmitter. In
particular, we recognize that due to the \changes{one-bit} DAC, precoding needs to
be done on a symbol-by-symbol basis. Moreover, we observe that it is
crucial to design the quadrature amplitude modulation (QAM)
constellation range for each given channel realization in order to
perform low-resolution \changes{transmit precoding} for each
symbol of the designed QAM constellation with reasonable accuracy. 
This paper focuses on the massive MIMO regime and uses the average
symbol error rate (SER) of uncoded transmission at the receiver as the design criterion. We
make the following main technical contributions toward the goal of
optimizing the {\it constellation range} and subsequently {\it one-bit
symbol-level precoding} for the massive MIMO downlink: 

\emph{Tight SER Expression:} A tight expression for the SER for QAM
transmission under one-bit symbol-level precoding is derived. 
Due to the low-resolution transmit DAC, the noiseless receive signal may
not be exactly at the intended constellation location. The derived expression
suggests that 
there is a reduction in the minimum constellation symbol distance in
this case. The minimum distance
is effectively reduced by the twice of the distance between the noiseless
received signal and the intended constellation symbol.

\emph{QAM Constellation Range Design:} We point out the importance of
optimizing the QAM constellation range with one-bit symbol-level precoding. 
Inspired by the results from the case where the infinite-resolution
DACs are available, we conclude that the constellation range should be
restricted to the regime where the noiseless received signal of
\changes{one-bit} precoding can approach all the points in the designed
constellation. Our analytical result shows that a reasonable choice
for the QAM constellation range for \changes{one-bit} precoding should be about
\changett{$\sqrt{2/\pi}$} times the optimal constellation range in the infinite-resolution
case with \changes{instantaneous (i.e., per-symbol)} power constraint \changem{in the massive MIMO limit.} 

\emph{Low-Complexity Algorithm for Precoding Design:} 
In \changes{one-bit} precoding, the transmitting signal of each antenna is selected
from a quadrature phase shift keying (QPSK) alphabet and hence the
transmitting signal design problem is a combinatorial
optimization problem for which exhaustive search would have
exponential complexity. This paper proposes a low-complexity two-step heuristic algorithm
which finds a high-quality solution for the \changes{one-bit} symbol-level
precoding problem. In the first step, the algorithm iteratively designs the transmitted signal
at each antenna such that the overall \changes{residual} signal 
is as close to the origin of the complex
plane as possible. This process is repeated until the transmit signals 
of all the antennas except last few, i.e., about $5-10$ antennas, are 
designed. In the second step,
the algorithm performs an exhaustive search on the \changes{transmit}
signals of the remaining antennas in order to find the best
match to the intended constellation point. The complexity of the proposed algorithm scales with
the square of the total number of antennas.  

\emph{Multi-User One-Bit Precoding:} The proposed constellation range 
and one-bit precoder designs are generalized to the multi-user case.
We first design the constellation range for the multi-user
zero-forcing precoder with per-symbol power constraint but assuming
infinite-resolution DACs. We observe that in the limit of large number 
of users, the constellation range for the multi-user case should be
set to be that of the single-user case multiplied by a factor which is
a function of the number of constellation points and the number of users. 
For the one-bit precoding case, we propose to further scale the
proposed constellation range design by \changett{$\sqrt{2/\pi}$}, 
\changem{the same factor as for the single-user massive MIMO.}
 The low-complexity precoding design
algorithm is likewise generalized to the multi-user scenario.

\emph{Performance Characterization:} 
We analytically characterize the performance of the proposed one-bit
symbol-level precoding design and show that \changes{in the massive MIMO regime} there is about $2$dB gap 
between the proposed \changes{one-bit} precoding scheme and infinite-resolution 
precoding (under the same per-symbol power constraint). 
This $2$dB gap is true for both single-user and multi-user
precoding designs. Moreover, we show that this $2$dB gap
translates to requiring about $50\%$ more antennas for
the \changes{one-bit} precoding architecture as compared to the conventional
infinite-resolution architecture for achieving the same performance. 

%%%%%%%%%%%%%%%%%%%%%%%%%%%%%%%%%%%%
%% I-B) Related Work
%%%%%%%%%%%%%%%%%%%%%%%%%%%%%%%%%%%%
\subsection{Related Work}
Most of the existing literatures on \changes{one-bit} beamforming focuses on the
uplink scenario in which the BS is equipped with \changes{one-bit}
analog-to-digital converters (ADCs). The performance analyses of such
systems for narrowband channels are provided in
\cite{risi2014massive,heath7155570,swindle7931630,studer7894211} while
those for wideband channels are presented in
\cite{Heath7600443,Durisi7458830}. 

\changes{ The early works on downlink precoding with one-bit DACs propose to use linear-quantized
precoding schemes in which the precoder is obtained by quantizing the traditional linear precoders
\cite{Nossek_2009,swindle7569670,swindle7946265}.  
\changes{Although} the class of linear-quantized precoding can approach the performance of the
infinite-resolution precoding in \changes{the} low signal-to-noise ratio (SNR)
regime, it suffers from a high symbol error floor in the high SNR regime.} \changet{In \cite{swidle7953405}, the authors propose some methods for slightly perturbing the transmitted signal of the quantized ZF scheme in \changes{\cite{swindle7569670,swindle7946265}} in order to improve performance at higher SNRs and numerically show that those perturbations can provide significant gains for QPSK signaling.}    
More sophisticated \changes{non-linear} precoding methods have been proposed in
\changes{\cite{jedda2016minimum,Nossek7472304,Usman7996519,Landau_BB_2017}} for the scenario
where both the receivers and the transmitter are equipped with \changes{one-bit}
ADCs and DACs, respectively. However, these methods are applicable
only for QPSK signaling and for the case where the receiver has
\changes{one-bit} ADC. We observe that in practice the user equipment is often 
equipped with relatively few number of antennas, so the use of 
high-resolution ADCs at the downlink receiver is 
often acceptable from a power consumption point of view.
For this reason, this paper considers the scenario with \changes{one-bit} DAC at
the transmitter, but high-resolution ADC at the receiver.

The authors of \cite{amor201716} generalize the algorithm in
\cite{jedda2016minimum} for the infinite-resolution receivers and
introduce a technique to transmit $16$-QAM symbols  with \changes{one-bit}
transmitters based on the idea of superposition coding. Unlike the
algorithm in \cite{amor201716}, the proposed scheme in this paper can
be applied to the QAM constellations with any size. More importantly,
this paper points out the crucial role played by QAM constellation
range design, which is not considered in prior work
\cite{amor201716}. 

The recent contributions of \cite{studer7952800,studer7967843} 
consider designing the constellation range, as well as a non-linear
precoder scheme with higher order modulations. The algorithms in
\cite{studer7952800,studer7967843}, which seek to minimize the mean
squared error at the user side, achieve excellent performance within 
few dB gap to the infinite-resolution precoding benchmarks.  However,
in the algorithm of \cite{studer7952800,studer7967843}, the
constellation range is designed for each symbol transmission, which means
that the range information needs to be communicated to the receiver as
side information on a symbol-by-symbol basis. This is a significant
overhead. \changet{The recent work \cite{studer7869149} extends the algorithm in \cite{studer7967843} by considering the constellation range design for $T$ symbol transmissions in which the channel remains constant. However, the high computational complexity of the algorithm in \cite{studer7869149} (which is $O(M^3T^3)$ for a system with $M$ transmit antennas) prohibits its use for the typical massive MIMO systems in which $M$ and $T$ are both large.}

\changet{By recognizing that the coherence time of the
wireless channel in practice is typically large enough to allow
at least 
several hundreds of symbol transmissions, this paper proposes to design the constellation range for each fading block such that it is suitable for all possible symbol vector choices from the constellation.}
Furthermore, this paper shows that for large-scale antenna
arrays the proposed constellation range design can be approximated by a constant
value which is independent of the channel realization thanks to the
``channel hardening'' phenomenon in massive MIMO systems
\cite{Marzetta1327795}. Therefore, the QAM constellation range
can be predetermined a priori with no need for instantaneous channel
state information (CSI) at the receivers. 

\changes{It is worth mentioning that the factor $\sqrt{2/\pi}$, which we propose as a constellation range reduction factor for one-bit precoding as compared to the infinite-resolution precoding scenario, also appears in the results of \cite{swindle7569670,studer7967843}, but for a different scheme of linear-quantized precoding, and using a completely different justification.}

\changes{Finally, one-bit precoding for signal constellations other than QAM constellation has also been considered, e.g., \cite{jedda2017psk,jedda2017massive} propose novel one-bit precoding designs for the phase shift keying (PSK) constellation. Comparing the performance of different constellation choices and finding the best constellation can be considered as an interesting direction for future work.}

%%%%%%%%%%%%%%%%%%%%%%%%%%%%%%%%%%%%
%% I-C) Paper Organization and Notations
%%%%%%%%%%%%%%%%%%%%%%%%%%%%%%%%%%%%
\subsection{Paper Organization and Notations}
The remainder of this paper is organized as follows.
Section~\ref{sec:sym} introduces the system model and the problem
formulation for massive MIMO \changes{one-bit} precoding. Section~\ref{sec:SER}
is devoted to deriving the SER expression in \changes{one-bit} precoding
architecture. For the single-user case, Section~\ref{sec:const_single}
considers designing the QAM constellation range while
Section~\ref{sec:prec_single} considers designing the precoder for the
fixed constellation range. Those designs are further generalized to
the multi-user case in Section~\ref{sec:multi}, where the performance
gap between the proposed \changes{one-bit} precoding design and the
infinite-resolution ZF precoder is also analyzed. Simulation results
are provided in Section~\ref{sec:sim} and conclusions are drawn in
Section~\ref{sec:con}.

This paper uses lower-case letters for scalars, lower-case bold face
letters for vectors and upper-case bold face letters for matrices. The
real part of a complex scalar $s$ is denoted by
$\operatorname{Re}\{s\}$. Further, we use the superscript ${}^H$ to
denote the Hermitian transpose of a matrix and superscript
${}^\dagger$ to denote the complex conjugate.  The identity and
all-one matrices with appropriate dimensions are denoted by
$\mathbf{I}$ and $\mathbf{1}$, respectively; $\mathbb{C}^{m\times n}$
denotes an $m$ by $n$ dimensional complex space;
$\mathcal{CN}(\mathbf{0},\mathbf{R})$ represents the zero-mean
circularly symmetric complex
Gaussian distribution with covariance matrix $\mathbf{R}$;
$\mathcal{N}(\boldsymbol{\mu},\mathbf{\Sigma})$ represents
a real Gaussian distribution with mean $\mathbf{\boldsymbol{\mu}}$
and covariance matrix $\mathbf{\Sigma}$. The notations
$\operatorname{Tr}(\cdot)$, $\operatorname{log}_{10}(\cdot)$, and
$\mathbb{E}\{\cdot\} $ represent the trace, decimal logarithm, and
expectation operators, respectively. Finally, $|\cdot|$ represents the
absolute value of a scaler while $\|\cdot\|_p$ indicates the $p$-norm
of a vector.

%%%%%%%%%%%%%%%%%%%%%%%%%%%%%%%%%%%
%Figure 1
%%%%%%%%%%%%%%%%%%%%%%%%%%%%%%%%%%%
\begin{figure}[t]
	\centering
	{\includegraphics[width=0.45\textwidth]{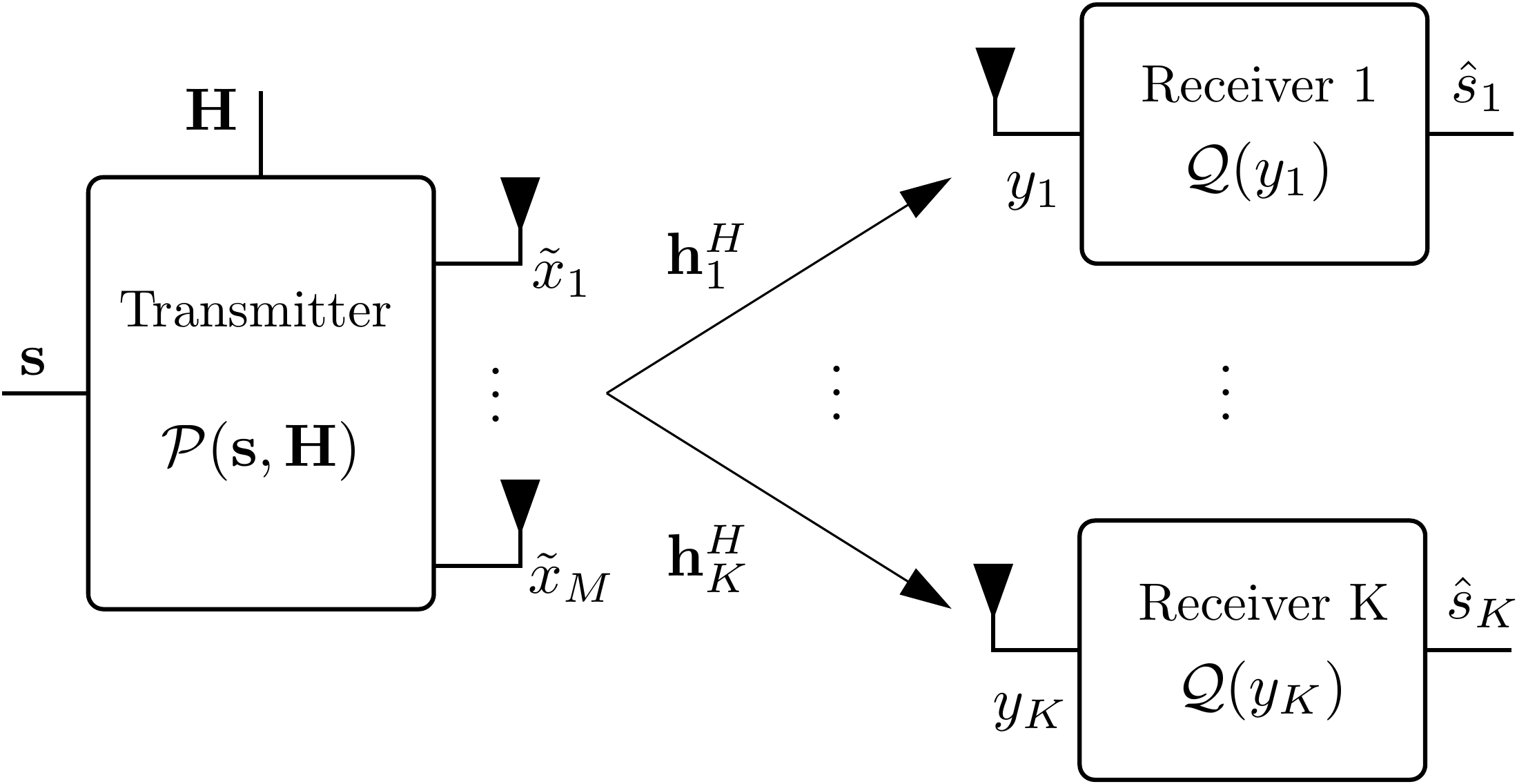}}
	\caption{A multi-user MISO system with symbol-level precoding.}
	\label{fig:sys_model}
\end{figure}

%%%%%%%%%%%%%%%%%%%%%%%%%%%%%%%%%%%
% II) System Model
%%%%%%%%%%%%%%%%%%%%%%%%%%%%%%%%%%%
\section{System Model}
\label{sec:sym}
Consider the downlink of a multi-user MISO system in which a BS with a
large number of antennas, $M$, serves $K$ single-antenna users. For
such a system, the received signal at user $k$ can be \changes{modeled} as
%1
\begin{equation}
y_k= \sqrt{\tfrac{P}{M}} \bh_k^H\bx+z_k,
\end{equation}
where \changes{$\bh_k^H \in \mathbb{C}^{1\times M}$ is the vector of channel gains} to user $k$, $\bx \in \mathbb{C}^{M\times 1}$ is the \changes{normalized} transmitted signal, $z_k \sim \mathcal{CN}(0,2\sigma^2)$ is the additive white Gaussian noise, and $P$ is the total transmit power budget. 

This paper seeks to design the transmitted signal when one-bit DACs
are employed at each antenna element. This means that the \changes{normalized}
transmitted signal $\bx$ must come from a finite alphabet, i.e.,
$\bx\in\X^M,$ where \changes{$\X=\left\{\tfrac{1}{\sqrt{2}}(\pm1\pm
\imath)\right\}$}, where $\imath$ is the imaginary unit satisfying
$\imath^2=-1$. In order to focus on the impact of \changes{one-bit} precoding,
this paper assumes that \changes{the full CSI is available at the transmitter.}

In conventional infinite-resolution precoding, the transmitted signal
is designed as a product of a beamforming vector and the symbol
constellation point. This is not possible to do when the transmitted
signal must come from a low-resolution alphabet.  This paper considers
instead a symbol-level precoding scheme, in which the BS designs 
the $M$-dimensional precoded transmitted signal, \changes{$\tilde{\bx} \triangleq \sqrt{\tfrac{P}{M}} \bx $}, directly on a symbol-by-symbol basis as a function of the instantaneous channel
state information (CSI), $\bH = [\bh_1,\ldots,\bh_K]^H$, and the
intended constellation point, $\bs = [s_1,\ldots,s_K]$, as
%2
\begin{equation}
\changes{\tilde{\bx} =\mathcal{P}\left(\bs,\bH\right),}
\end{equation}
where the function $\mathcal{P}: {\mathbb{C}^M}\times \mathbb{C}^{K\times M}\rightarrow \mathcal{X}^M$ represents the precoder. 

\changet{At the receiver side, we do not assume the availability of the CSI, and instead assume that 
in each coherence \changes{block} of the channel, 
the constellation range and the constellation size are fed 
  back to the users such that each user seeks to recover its intended symbol, $s_k$, from its received signal, $y_k$, by mapping $y_k$ to its nearest constellation point, i.e., $\hat{s}_k = \mathcal{Q}(y_k)$. 
The overall system model is shown in Fig.~\ref{fig:sys_model}.
\changes{Note that the concept of symbol-level precoding is considered in the
earlier works \cite{Joham909609,Masouros4801492,Alodeh7042789,alodeh7942010,Alodeh7404291,Masouros8023970,Masouros7103338} for the infinite-resolution 
case---it is adopted for the finite-resolution DAC context considered
in this paper.}} 
\changem{We note here that symbol-level precoding requires transmit beamforming adaptation at the symbol rate, rather than at the timescale of channel coherence time as in the case of traditional beamforming. This translates to significantly more demanding transmit processing speed requirement.}

\begin{figure}[t]
	\centering
	{\includegraphics[width=0.37\textwidth]{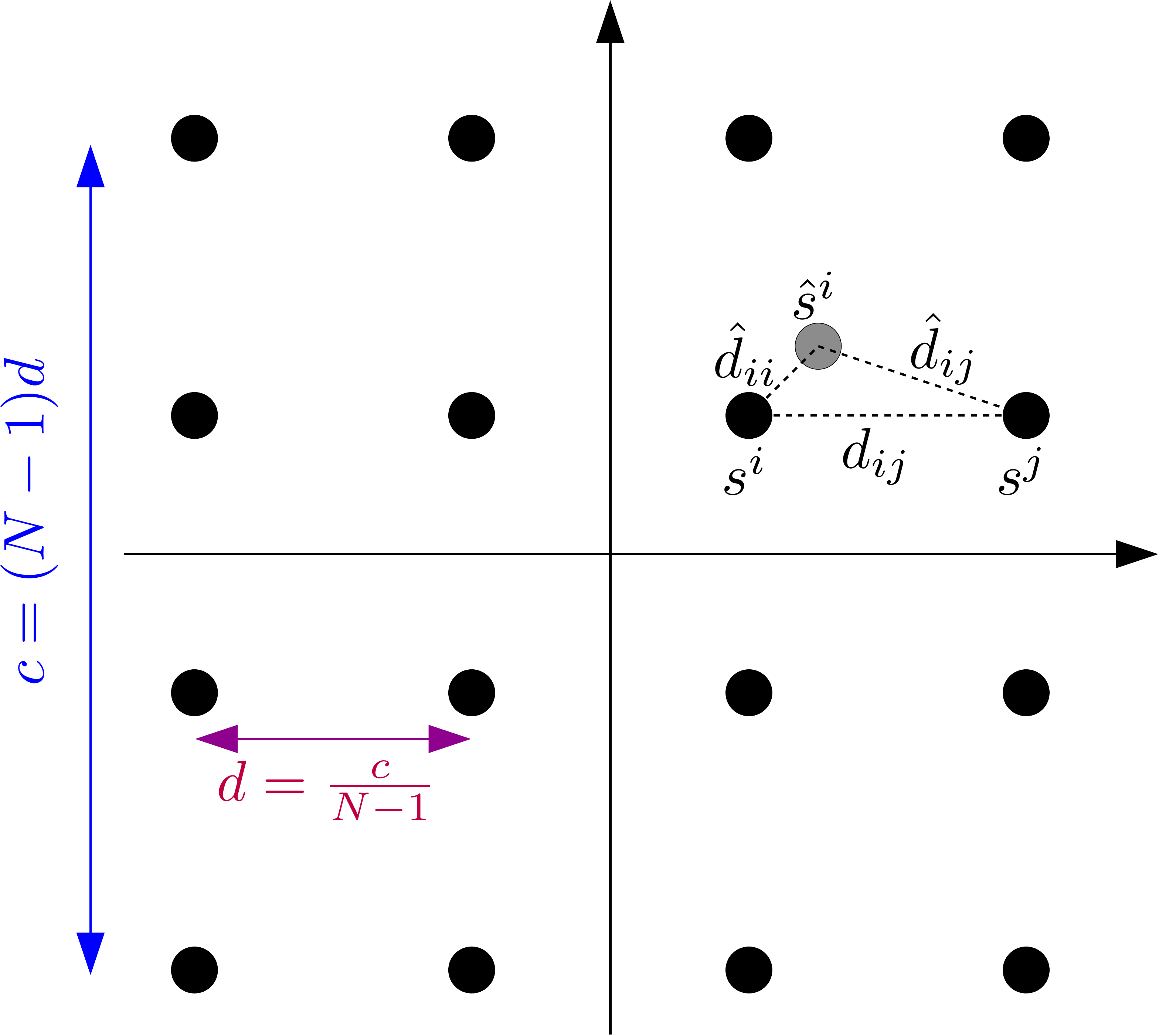}}
	\caption{An example of $N^2$-QAM constellation for $N=4$.}
	\label{fig:QAM}
\end{figure}

This paper restricts attention to uncoded square QAM constellations and also
explicitly designs the constellation range for one-bit symbol-level
precoding. In particular, we use $N^2$-QAM constellations with
the range of $c$ in each dimension and the minimum distance of $d =
\tfrac{c}{N-1}$; an example for $N = 4$ is depicted in
Fig.~\ref{fig:QAM}.  We assume a block-fading channel model in
which the channel stays constant for at least hundreds of symbol
transmissions so that the constellation range $c$ and the constellation
size $N$ only need to be communicated to the receiver as preamble on a 
per-fading-block basis. \changem{It is emphasized that this paper considers the adaptation of the constellation range to the channel state information only, while the number of constellation points are assumed to be fixed. A complete adaptive modulation scheme in which both the constellation range and the number of constellation points are jointly optimized can be considered as future work.
We also emphasize that the techniques presented in this paper pertain to uncoded transmission only. A complete characterization of the capacity and the optimal transmission strategy for the finite-resolution-input massive MIMO channel is still an open problem.} 

\changet{In general, for the multi-user scenario, different constellations can be designed for different users. However, similar to the references \cite{studer7952800,studer7967843,studer7869149}, this paper considers a simplified problem in which the symbols of
all users come from the same constellation.} 
This assumption may force the users with strong channels to operate
below their actual capacities, but such fairness concern can be
addressed by an intelligent scheduler that groups
users with \changes{nearly} similar channels within each
time-frequency slot. For this reason, the rest of the paper assumes
that all users have the same large-scale fading, i.e., $\bh_k \sim
\mathcal{CN}(\mathbf{0},\bI),$ $\forall k$.

The problem of interest is to design the QAM constellation range in each coherence time of the channel, and then design the transmitted signal corresponding to each symbol vector where each element of that symbol vector is chosen from the designed constellation in order to minimize the average symbol error rate.  \changet{The proposed constellation range design for the multi-user scenario is based on the i.i.d Rayleigh fading channel model
in which a similar large-scale fading is assumed for all the users. Generalizing this design for other channel models
require further efforts and it can be considered as future work.}

%%%%%%%%%%%%%%%%%%%%%%%%%%%%%%%%%%%
%Figure 2
%%%%%%%%%%%%%%%%%%%%%%%%%%%%%%%%%%%

%%%%%%%%%%%%%%%?%%%%%%%%%%%%%%%%%%%%%
%% III) SER Characterization for Symbol-Level Precoding
%%%%%%%%%%%%%%%%%%%%%%%%%%%%%%%%%%%%
\section{SER Characterization for Symbol-Level Precoding}
\label{sec:SER}
This section provides an SER expression of a generic user in a
multi-user MISO system with one-bit symbol-level precoding
described in Section~\ref{sec:sym}. In order to avoid encumbering the
notation, we drop the user index, $k$, in the notations throughout
this section. 

Let us assume that the intended constellation point for the considered
receiver is $s^i$. Due to the finite-resolution DAC constraint in
one-bit precoding, the noiseless received signal, $\hat{s}^{i} =
\sqrt{\tfrac{P}{M}} \bh^H\bx_i$, may not exactly coincide with the intended
constellation point. \changet{In this case, the received signal is $y = \hat{s}^i + z$ and
the \changet{pairwise error probability},
$\operatorname{Pr} \left(s^i \rightarrow
s^j\right)$, can be written as
%3
\begin{eqnarray}\nonumber
\operatorname{Pr}\left(s^i \rightarrow s^j\right) &\overset{\bigtriangleup}{=}& \operatorname{Pr}\left(\left| y - s^i \right| \geq \left| y - s^j \right| \right) \\ \nonumber
&\overset{(a)}{=}& \operatorname{Pr}\left(\left| \hat{s}^i + z - s^i \right|^2 \geq \left| \hat{s}^i + z - s^j \right|^2\right) \\ \nonumber
&\overset{(b)}{=}& \changet{\operatorname{Pr}\left(\operatorname{Re}\{\tilde{z}\} \geq \frac{\left| \hat{s}^i - s^j \right|^2 - \left| \hat{s}^i - s^i \right|^2}{2 \left| s^i - s^j \right|}\right) \label{4.7.d}}\\ 
&\overset{(c)}{=}&  Q\left(\frac{\hat{d}_{ij}^2 - \hat{d}_{ii}^2}{2d_{ij}\sigma} \right),
\label{SERSiSj_1}
\end{eqnarray}
where $Q(u) = \tfrac{1}{\sqrt{2\pi}} \int_u^\infty e^{ - \tfrac{v^2}{2}} dv$, 
\changet{$\tilde{z} = ze^{\imath\angle \left(s^j - s^i\right)^\dagger}$ is a Gaussian random variable with the same distribution as $z$, i.e., $\tilde{z}\sim \mathcal{CN}\left(0,2\sigma^2\right)$},
$\hat{d}_{ij} \overset{\bigtriangleup}{=} \left| \hat{s}^i - s^j \right|$, and $d_{ij} \overset{\bigtriangleup}{=} \left| {s}^i - s^j \right|$. 
\changet{
In the above equations, $(a)$ is obtained by substituting $y = \hat{s}^i + z$,  $(b)$ is obtained by defining $\tilde{z} = ze^{\imath\angle \left(s^j - s^i\right)^\dagger}$ and rearranging the inequality for $\tilde{z}$, and $(c)$ is based on the definition of the $Q$-function}. }

\changet{Similar to the conventional SER analysis for a reasonable
SNR range \cite{Proakis2008}, the pairwise probability error of the closest \changes{neighboring} constellation points in \eqref{SERSiSj_1} can be used to tightly approximate the overall SER as
%4
\begin{equation}
\label{SER_approx_new}
\operatorname{SER} \approx \frac{1}{N^2} \sum_i \sum_{j\in\bar{\mathcal{N}}_i} Q\left(\frac{\hat{d}_{ij}^2 - \hat{d}_{ii}^2}{2d_{ij}\sigma} \right),
\end{equation}
where $\bar{\mathcal{N}}_i$ is the set of nearest constellation points to $s^i$.}

The expression in {\eqref{SER_approx_new}} is complicated and is difficult to use
as a metric to design the constellation range. As a
result, we seek to find an upper-bound for the expression in
{\eqref{SER_approx_new}}.  As it can be seen from Fig.~\ref{fig:QAM},
$\hat{d}_{ii}$ , $\hat{d}_{ij}$, and ${d}_{ij}$ can be considered as
three edges of a \changes{triangle}, and hence by using the triangular
inequality, we can write 
%5
\begin{eqnarray}\nonumber
\frac{\hat{d}_{ij}^2 - \hat{d}_{ii}^2}{2d_{ij}} &=& \frac{(\hat{d}_{ij}+\hat{d}_{ii})(\hat{d}_{ij}-\hat{d}_{ii})}{2d_{ij}}  \\
&\geq& \frac{({d}_{ij})({d}_{ij}-2\hat{d}_{ii})}{2d_{ij}}\hspace{10pt} = \hspace{6pt} \frac{{d}_{ij}-2\hat{d}_{ii}}{2}. 
\end{eqnarray}

\changet{Now, using that $Q(\cdot)$ is a decreasing function and $d_{ij} = d$ for all $j\in\bar{\mathcal{N}}_i$ where $d$ denotes the minimum distance in the considered QAM constellation, we can write the following upper-bound on the SER expression in \eqref{SER_approx_new} as:
%6
\begin{equation}
\operatorname{SER} \lesssim \sum_i \frac{g_i}{N^2} Q\left(\frac{d-2\hat{d}_{ii}}{2\sigma}\right)
\label{SERu_final}
\end{equation}
where $g_i$ is the number of \changes{minimum-distance neighbors} of the symbol $s^i$.}

The rest of the paper first considers designing the constellation range and the
non-linear precoder for the single-user scenario, $K=1$, by minimizing
the expression \eqref{SERu_final}, then 
generalizes the proposed design to the multi-user case.
 
%%%%%%%%%%%%%%%?%%%%%%%%%%%%%%%%%%%%%
%% IV) Constellation Range Design for Single-User Scenario
%%%%%%%%%%%%%%%%%%%%%%%%%%%%%%%%%%%%
\section{Constellation Range Design for Single-User Scenario}
\label{sec:const_single}
This section proposes an appropriate choice of constellation range, $c$, 
for \changes{one-bit symbol-level precoding} under a fixed MISO channel serving a
single user.
In order to gain insight on how to design $c$ for \changes{one-bit}
precoding, we first consider designing the constellation range
of symbol-level precoding with infinite-resolution DACs under
\changes{instantaneous per-symbol total power constraint} across the antennas, 
and also under \changes{instantaneous per-symbol} per-antenna power constraint. The instantaneous
per-symbol power constraint refers to that each precoded symbol needs
to satisfy a power constraint (instead of the average power across
multiple symbols.)

%%%%%%%%%%%%%%%?%%%%%%%%%%%%%%%%%%%%%
%% IV.A) Infinite-Resolution Precoding with Total Power Constraint
%%%%%%%%%%%%%%%%%%%%%%%%%%%%%%%%%%%%
\subsection{Infinite-Resolution Precoding with Total Power Constraint}
\label{subsec_inf_t}
Under symbol-level precoding, the transmit signal is designed as a
function of the constellation point and the instantaneous channel
realization. The range of the QAM constellation is an important
parameter that needs to be designed to optimize performance.
Assuming \changes{a square} $N^2$-QAM and using the expression \eqref{SERu_final} as
the metric for minimizing the SER with $d = \tfrac{c}{N-1}$ and 
$\hat{d}_{ii} = \left| \sqrt{\tfrac{P}{M}} \bh^H\bx_i - s^i \right|$,
the problem of optimizing the constellation range for symbol-level
precoding with infinite-resolution precoding and with 
\changes{instantaneous per-symbol} total power constraint can be written as
%7
\begin{eqnarray}
\min_{c \geq 0} \sum_{i=1}^{N^2} \frac{g_i}{N^2} \min_{\|\bx_i\|^2_2 \leq M} 
Q\left(\frac{\tfrac{c}{N-1}-2 \left| \sqrt{\tfrac{P}{M}} \bh^H\bx_i - s^i \right|}{2\sigma}\right).
\label{prob_inf_pt}
\end{eqnarray}

With infinite resolution on ${\bf x}_i$, the optimal solution ${\bf x}_i$ 
\changes{for the inner minimization problem of \eqref{prob_inf_pt} with fixed $c$} is
%8
\begin{equation}
\bx_i =  \frac{\sqrt{M}\mathbf{h}}{\| \mathbf{h}\|_2}\min{\left\{ 1, \frac {|s^i|}{\sqrt{P}\|\mathbf{h}\|_2} \right\}} e^{\imath\angle{s^i}}.
\end{equation}

Intuitively, the optimal ${\bf x}_i$ is to match the channel. With
infinite resolution on ${\bf x}_i$, the precoded signal can
reconstruct $s^i$ perfectly as long as 
%9
\begin{equation}
|s^i| \le \sqrt{P}\|\mathbf{h}\|_2.
\end{equation}

By substituting this solution for all $\bx_i$ in
\eqref{prob_inf_pt} and taking some simple algebraic steps, the
problem \eqref{prob_inf_pt} for designing $c$ can be rewritten as
%10
\begin{eqnarray}
\min_{c \geq 0} \sum_i \frac{g_i}{N^2} Q\left(\frac{\tfrac{c}{N-1}-2 \max\left\{ 0, \left| s^i\right| - \sqrt{P} \| \mathbf{h}\|_2 \right\} }{2\sigma}\right).
\label{prob_inf_pt_onlyc}
\end{eqnarray}

In \eqref{prob_inf_pt_onlyc}, the Q-function can be \changes{approximated} by $Q(x) \approx \tfrac{1}{12} e^{- \tfrac{x^2}{2} } + \tfrac{1}{4} e^{-\tfrac{2}{3} x^2},~ x> 0$\cite{Simon1210748}. This exponential relation between the output of the $Q$-function and its argument suggests that the value of the objective function in \eqref{prob_inf_pt_onlyc} is dominated by the error probability of symbols which result in the smallest argument of the $Q$-function. Further, since $Q(\cdot)$ is a decreasing function, the smallest argument occurs for the furthest constellation points which have the maximum $\left|s^i\right|$, i.e., $\max_i \left|s^i\right|= \tfrac{c}{\sqrt{2}}$. Therefore, it is desirable to consider designing the constellation range such that the argument of the $Q$-function for those furthest constellation points is maximized, i.e.,
%11
\begin{equation}
\max_{c \geq 0} \left\{ \frac{c}{N-1}-2 \max\left\{ 0, \frac{c}{\sqrt{2}} - \sqrt{P} \| \mathbf{h}\|_2\right\} \right\}.
\label{c_design}
\end{equation}

The objective function of \eqref{c_design} is a piecewise linear function of $c$ and the optimal solution \changes{for} that, $c^{*}_{\text{inf,t}}$, can be calculated as
%12
\begin{equation}
c^{*}_{\text{inf,t}} = \sqrt{2P} \| \bh\|_2.
\label{c_inf_t}
\end{equation}
This result has the following interpretation. As mentioned earlier,
with 
infinite-resolution symbol-level precoding under instantaneous total power
constraint, we can construct any point in the complex plane inside a
circle \changes{centered} at the origin of the complex plane with radius
$\sqrt{P} \| \bh\|_2$ exactly as the noiseless received signal. The
$\sqrt{2}$ factor comes from the fact that 
the distance of the furthest constellation point from the origin for a square constellation is
$\tfrac{c}{\sqrt{2}}$. The expression in \eqref{c_inf_t} suggests that
we should increase the range of the constellation, $c$, as much as
possible subject to the constraint that the furthest constellation
point can be constructed by symbol-level precoding. In other words,
the furthest constellation points should be designed to be at the
distance of $\sqrt{P} \|\bh\|_2$ from the origin.

The constellation range in \eqref{c_inf_t} is a function \changem{of} the channel
realization, $\bh$. This means that $c^{*}_{\text{inf,t}}$ should
adapt to the realization of the channel in each coherence time. 
However, for large-scale antenna arrays, it is possible to show that $
\| \bh\|_2$ can be well approximated by 
 a constant $\sqrt{M}$.
This phenomenon, called channel hardening in massive MIMO literature \cite{Marzetta1327795}, suggests that for large $M$ we can approximate 
%13
\begin{equation}
c^{*}_{\text{inf,t}} \overset{M\to \infty}{\approx} \sqrt{2PM},
\label{c_inf_t_M}
\end{equation}
without significant performance degradation. By this design, the
constellation range is a function of the path-loss only and not
specific Rayleigh fading component of ${\bf h}$. 

%%%%%%%%%%%%%%%?%%%%%%%%%%%%%%%%%%%%%
%% IV.B)  Infinite-Resolution Precoding with Per-Antenna Power Constraint
%%%%%%%%%%%%%%%%%%%%%%%%%%%%%%%%%%%%
\subsection{Infinite-Resolution Precoding with Per-Antenna Power Constraint}
This paper eventually considers a one-bit precoder for which every 
antenna has the same transmit power, so as an intermediate step, it
is worth considering infinite-resolution precoding with
per-antenna power constraint rather than total power constraint. In
this case, the \changes{normalized} transmitted signal of each antenna should satisfy
$|x_m| \leq 1$.

Fixing $\bh$, we claim that the complex numbers that can
be realized with $ \sqrt{\tfrac{P}{M}} \bh^H\bx$ under the constraints
$|x_m| \leq 1,~\forall m$, are exactly all the points inside a circle
centered at the origin of the complex plane, but with a reduced radius
as compared to the total power constraint case. To show this, let us consider the following optimization problem: 
%14
\begin{equation}
\min_{|x_m| \leq 1, ~\forall m} ~ \left|\sqrt{\tfrac{P}{M}} \bh^H\bx - s \right|.
\label{x_m_problem}
\end{equation}

The optimal solution for $\bx$ in \eqref{x_m_problem} is given as
%15
\begin{equation}
x_m = \min \left\{\sqrt{\frac{M}{P\|\bh\|_1}} |s|,1\right\}
e^{\imath(\angle{h_m}+\angle{s})}, \quad \forall m,
\label{x_m}
\end{equation}
  where \changes{$h_m$ is the $m^\text{th}$ element of $\bh$.}
Now by substituting \eqref{x_m} into the objective function \changes{of} \eqref{x_m_problem}, we can see that only for complex numbers $s$ inside a circle of 
$|s| \leq \sqrt{\tfrac{P}{M}} \|\bh\|_1$, it is possible to realize
$s$ exactly. We denote the radius of this circle by
%16
 \begin{equation}
r^* =\sqrt{\frac{P}{M}} \|\bh\|_1.
\end{equation}
We remark that $r^*$ 
is also the largest range of $ \sqrt{\tfrac{P}{M}} \bh^H\bx$, i.e., 
%17
 \begin{equation}
r^* =\max_{|x_m| \leq 1, \forall m} {\left| \sqrt{\tfrac{P}{M}} \bh^H\bx \right|}.
\label{rstar_inf_p}
\end{equation}
This interpretation will be useful in the constellation range design 
for \changes{one-bit} precoding.

Using a similar argument as in Section~\ref{subsec_inf_t}, it can %now
 be shown that the optimal constellation range for minimizing the SER is
the largest constellation range such that all the constellation points
can be constructed accurately. Therefore, the furthest constellation
point, which is at the distance $\tfrac{c}{\sqrt{2}}$ for a square constellation, should be
at a distance $r^{*}$ from the origin, yielding that the optimal constellation
range \changes{is} %given by
%18
\begin{equation}
c^{*}_{\text{inf,p}}  = \sqrt{\frac{2P}{M}} \| \mathbf{h}\|_1.
\label{c_inf_p}
\end{equation}

\changet{For the considered channel model in which each element of the channel vector has an i.i.d Gaussian distribution, i.e., $h_j \sim \mathcal{CN}(0,1)$, we have $\mathbb{E}\left\{ |h_j| \right\} = \sqrt{\pi/4}$. Now, using the law of large number
in the large-scale antenna array limit when $M\to \infty$, we can write 
$ \tfrac{\|\bh \|_1}{M} \to \sqrt{\pi/4}$. Therefore, in systems with massive antenna arrays, it is possible to approximate $ \tfrac{\|\bh \|_1}{\sqrt{M}}$ as \changet{$\sqrt{(\pi/4)M}$}}. This result can be used to further simplify the constellation range expression in \eqref{c_inf_p} as 
%19
\begin{equation}
c^{*}_{\text{inf,p}}  \overset{M\to\infty}{\approx} \changet{\sqrt{\pi/4}}\sqrt{2PM}.
\label{c_inf_p_M}
\end{equation}

\changes{It} can be seen from \eqref{c_inf_t_M} and \eqref{c_inf_p_M} that
the ratio between the constellation ranges in infinite-resolution
precoding under total power constraint and under per-antenna power
constraint  in the massive MIMO limit converges to a constant:
%20
\begin{equation}
\frac{c^{*}_{\text{inf,p}}}{c^{*}_{\text{inf,t}}} \rightarrow \changet{\sqrt{\pi/4}},
\quad \mathrm{as} \quad  M\to \infty, 
\label{ratio_t_p}
\end{equation}
i.e., under the per-antenna power constraint, the constellation size
should be reduced by about 88\% in order for the symbol-level precoder
to be able to synthesize the constellation point.

%%%%%%%%%%%%%%%?%%%%%%%%%%%%%%%%%%%%%
%% IV.C) \changes{one-bit} Precoding
%%%%%%%%%%%%%%%%%%%%%%%%%%%%%%%%%%%%
\subsection{One-Bit Precoding}
\label{sec:1bit_range_single}
We now consider \changes{one-bit} symbol-level precoding where $\bx\in\X^M$ \changes{in which
$\X=\left\{\tfrac{1}{\sqrt{2}}(\pm1\pm \imath)\right\}$}. Unlike the
infinite-resolution case, the realizations of $\sqrt{\tfrac{P}{M}}
\bh^H\bx$ is no longer a continuum. Thus, it is no longer always true
that all constellation points can be constructed exactly at the
transmitter.  Nevertheless, as long as the reconstruction error is
sufficiently small (e.g., below the noise level), one-bit precoding
can still provide good performance.

\begin{figure}[t]
	\centering
	{\includegraphics[width=0.45\textwidth]{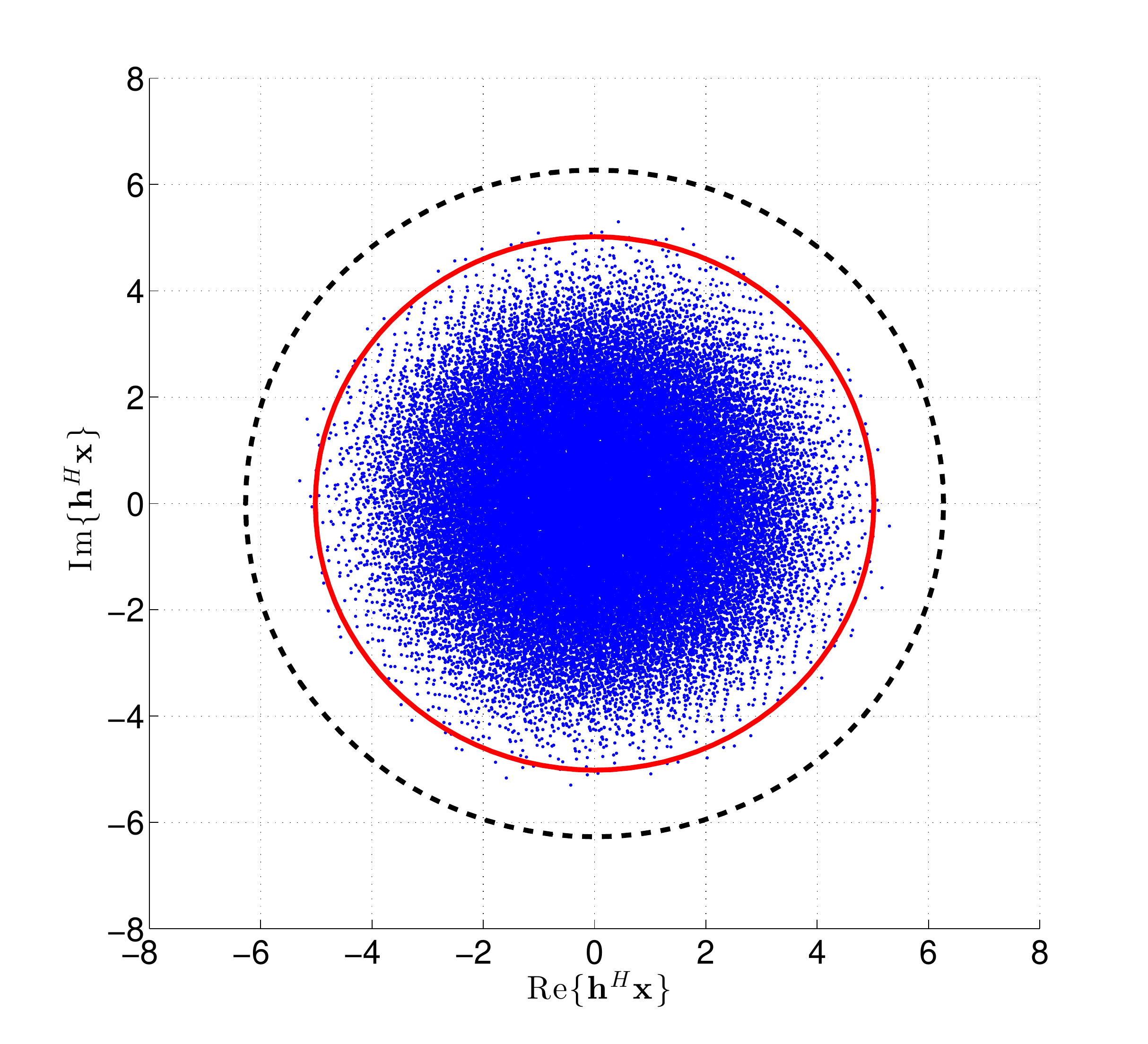}}
	\caption{Fix one random channel realization with $M=8$ and $P = 4$.
The data points are the $4^{M}$ possible realizations
of $\mathbf{h}^H\bx$, and the dashed black and solid red lines are
circles with radius of $c^{*}_{\text{inf,t}}  = \sqrt{2P} \|
\mathbf{h}\|_2$ and \changes{$\sqrt{2/\pi}c^{*}_{\text{inf,t}} $}, respectively. Most
data points are confined within the solid red circle.}
	\label{fig:scatter}
\end{figure}

To gain some intuition, Fig.~\ref{fig:scatter} is a scatter plot of
all $4^M$ possible realizations of $\bh^H\bx$ for
the case of $M=8$ antennas for a fixed $\bh$. It can be seen that all these points are confined
in a circle centered at the origin, yet the radius of the circle in
which these points concentrate is further reduced as compared to the
infinite-resolution cases discussed earlier.  

This paper aims to make a case that similar to the infinite-resolution case, i.e., \eqref{rstar_inf_p}, if the constellation range is
designed such that the furthest constellation point is located at
$\bar{r}^{*}$, where  
%21
\begin{equation}
\bar{r}^{*} = \max_{\bx \in \mathcal{X}^M} {\left| \sqrt{\tfrac{P}{M}} \bh^H\bx \right|},
\end{equation}
then all the constellation points can be reconstructed accurately.
This is clearly a {\it necessary} condition for constellation range.
But, as shown numerically later in the paper, when the number of transmit
antennas $M$ is large, this is also a {\it sufficient} condition,
i.e., an appropriate \changes{one-bit} precoding design can approach any complex
number inside the circle with radius $\bar{r}^{*}$ with negligible
error. 

Following this design strategy, the constellation range design problem
for \changes{one-bit} precoding can be written as
%22
\begin{equation}
c^{*}_{\text{$1$-bit}} = \sqrt{2} \max_{\bx \in \mathcal{X}^M} {\left| \sqrt{\tfrac{P}{M}} \bh^H\bx \right|}.
\end{equation}
The numerical evaluation of the above maximization is, unfortunately, difficult. 
Instead of solving this problem directly, this paper proposes to
reduce the constellation range for \changes{one-bit} precoding (which
automatically satisfies per-antenna power constraint) as compared to
infinite-resolution precoding with per-antenna power constraint by a
constant factor. The following result 
\cite{zhang2006complex} helps us identify such a factor. 

%%%%%%%%%%%%%%%%%%%%%%%%%%%%%%%%%%%
%Figure 3
%%%%%%%%%%%%%%%%%%%%%%%%%%%%%%%%%%%

Consider the following optimization problem
%23
\begin{subequations}
\begin{eqnarray}
&\displaystyle \max_\bx  &\bx^H \bA \bx\\ 
&\text{\normalfont{s.t.}} &x_m \in \mathcal{F}_q, \hspace{0.1in}\forall m,
\end{eqnarray}
\label{lemma_opt_1}
\end{subequations} 
where $\mathcal{F}_q = \left\{1,e^{\imath\tfrac{2\pi}{q}},\ldots,e^{\imath\tfrac{2\pi(q-1)}{q}} \right\}$.
Further, consider the following semidefinite programming (SDP) relaxation of \eqref{lemma_opt_1} which is now a convex optimization problem,
%24
\begin{subequations}
\begin{eqnarray}
 &\displaystyle{\max_{\bX \in \mathbb{C}^{M\times M}}}  &   \operatorname{Tr} (\bA\bX   )  \\
 &\text{\normalfont{s.t.}} & \mathbf{X} \succeq \mathbf{0},\\
 & & \mathbf{X}(i,i) = 1, \hspace{0.1in}\forall i=1,\dots,M.
\end{eqnarray}
\label{lemma_opt_SDP}
\end{subequations}
Suppose $\bX^{*}$ is an optimal solution of problem \eqref{lemma_opt_SDP}. To generate a feasible $\bx$ from the solution of  problem \eqref{lemma_opt_SDP}, draw a random vector $\boldsymbol{\zeta} \sim \mathcal{CN}(\mathbf{0},\bX^{*})$, and then quantize each element of $\boldsymbol{\zeta}$ to the nearest point in 
$\mathcal{F}_q$. The expectation of the objective function of the
solution obtained using the above randomization procedure can be shown
to be greater than $\alpha_q   \operatorname{Tr} ( \bA \bX^{*})$ \cite{zhang2006complex}, where
%25
\begin{equation}
 \alpha_{q} = \begin{cases} \frac{2}{\pi}, &\mbox{\normalfont{if} } q=2, \\
\frac{q^2(1-\cos{{\tfrac{2\pi}{q} }})} {8\pi}, & \mbox{\normalfont{if} } q\geq3.  \end{cases}
\end{equation}

For the given $q$, let us denote the objective function in
\eqref{lemma_opt_1} for the optimal solution as $f^{*}_q$ when $\bA =
\tfrac{P}{M}\mathbf{h}\mathbf{h}^H$. Then, the constellation range for
infinite-resolution precoding with per-antenna power constraint and
\changes{one-bit} precoding can be written as $c^{*}_{\text{inf,p}} =
\sqrt{2f^{*}_\infty}$ and $c^{*}_{\text{$1$-bit}} = \sqrt{2f^{*}_4}$,
respectively. \changet{Now since it is difficult to exactly characterize
$f^{*}_4$, we propose to approximate $\tfrac{f^{*}_4}{f^{*}_\infty}$
by the ratio of the SDP approximation bounds $\tfrac{\alpha_4}{\alpha_\infty}=\changet{\tfrac{2/\pi}{\pi/4} = 8/\pi^2}$. This means that 
%26
\begin{equation}
\label{c_1bit_single}
\changet{c^{*}_{\text{$1$-bit}} \approx \sqrt{8/\pi^2} c^{*}_{\text{inf,p}} \approx \sqrt{2/\pi} c^{*}_{\text{inf,t}}},
\end{equation}
where the last approximate equality follows from \eqref{ratio_t_p}. }

%%%%%%%%%%%%%%%%%%%%%%%%%%%%%%%%%%%
%Figure 4
%%%%%%%%%%%%%%%%%%%%%%%%%%%%%%%%%%%
\begin{figure}[t]
	\centering
	{\includegraphics[width=0.45\textwidth]{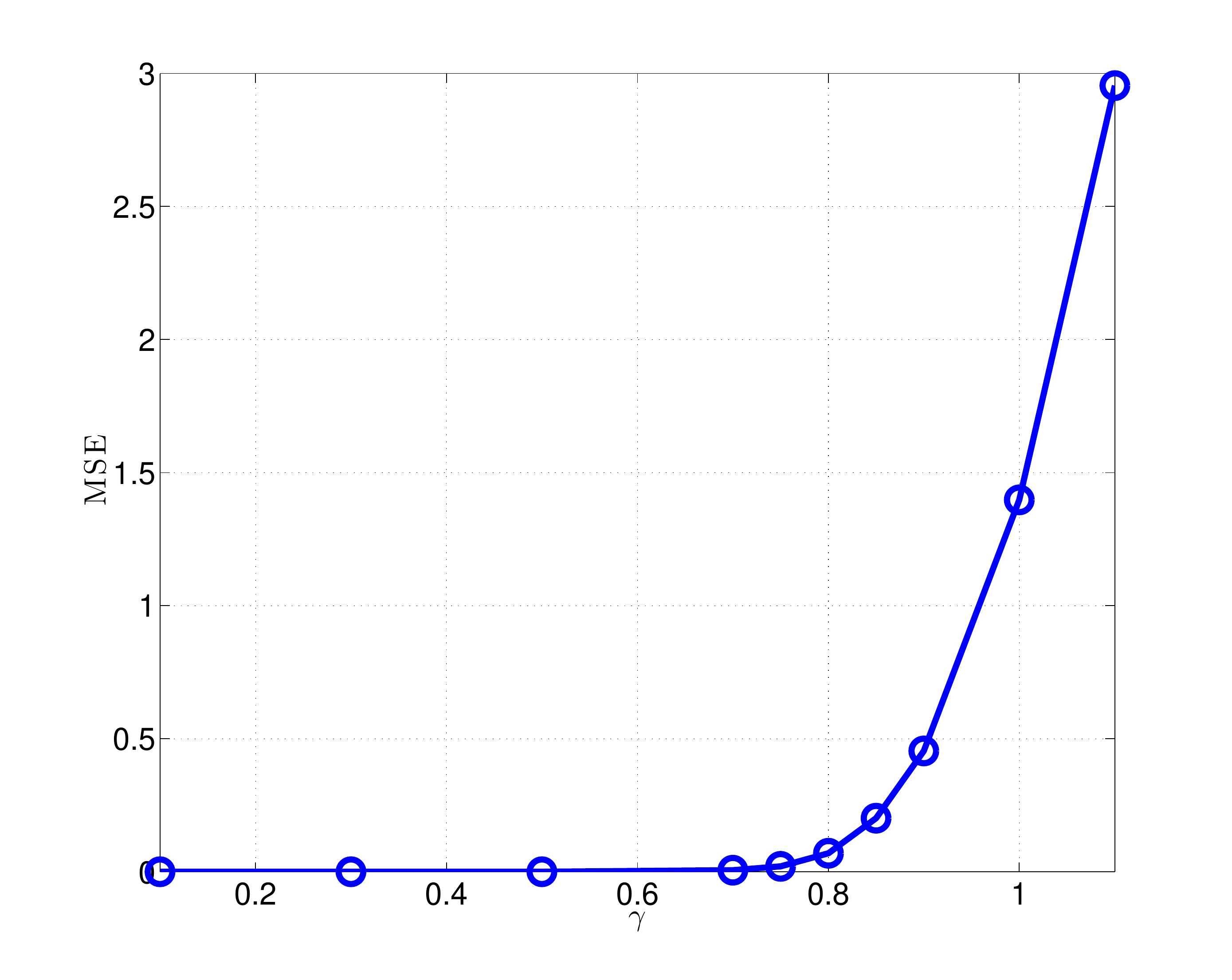}}
	\caption{The MSE versus $\gamma$ for a system with $M=8$ and $P=4$.
There is a phase transition at \changes{$\gamma_\text{th}=\sqrt{2/\pi} \approx 0.8 $.}}
	\label{fig:M8}
\end{figure} 

To see that the proposed constellation range is an appropriate choice,
let us consider as an example of a relatively small system with $M=8$
antennas and $P=4$. 
For a randomly generated fixed $\bh$,
Fig.~\ref{fig:scatter} plots all $4^M$ possible realizations of
$\bh^H\bx$. Fig.~\ref{fig:scatter} also plots two circles: a dashed
black circle with radius of $c^{*}_{\text{inf,t}}  =
\sqrt{2P}  \|\mathbf{h}\|_2$ and a solid red circle with the radius of
$\changett{\sqrt{2/\pi}}c^{*}_{\text{inf,t}}$ where the latter is the proposed value for
$c^{*}_{\text{$1$-bit}}$. It can be seen visually that almost all the
realizations of $\bh^H\bx $ are located inside the solid red circle
suggesting that the proposed design is a necessary condition for
constellation range.

To see that the proposed design is also a sufficient condition for constellation range, let us define the mean squared error (MSE) as 
%27
\begin{equation}
 \operatorname{MSE}(s) = \mathbb{E}_\bh\left\{\left| \sqrt{\tfrac{P}{M}}\bh^H\bx - s\right|^2\right\},
 \label{MSE_exp}
 \end{equation}
   and the parameter 
   %28
   \begin{equation}
   \label{gamma_definition}
   \gamma =  \tfrac{|s|}{\sqrt{P}\| \bh\|_2}.
   \end{equation}
Fig.~\ref{fig:M8} plots the average MSE against the parameter $\gamma$
over $10^{3}$ channel realizations.  The precoder $\mathbf{x} \in
\mathcal{X}^M$ is found by exhaustive
search in order to minimize the squared error for given $s$ and $\bh$.
It can be seen from  Fig.~\ref{fig:M8} that there is a phase transition 
around $\gamma_{\text{th}} =\changett{\sqrt{2/\pi} \approx 0.8}$ before which 
the MSE is very \changes{small,} while the MSE starts to increase
significantly after $\gamma_{\text{th}}$. This means that any complex
number inside the circle with radius  $c^{*}_{\text{$1$-bit}} =
\changett{\sqrt{2/\pi}}\sqrt{2P}  \|\mathbf{h}\|_2 $ can be realized \changes{accurately,} while it
is infeasible to do so for complex numbers outside of that circle,
thereby justifying the choice $c^{*}_{\text{$1$-bit}} = \changett{\sqrt{2/\pi}} c^{*}_{\text{inf,t}}$.
Further numerical evidence is provided in Section~\ref{sec:sim} to
support this choice under practical precoding and when $M$ is large. 

%%%%%%%%%%%%%%%?%%%%%%%%%%%%%%%%%%%%%
%% V) One-Bit Precoding Design for Single-User Scenario
%%%%%%%%%%%%%%%%%%%%%%%%%%%%%%%%%%%%
\section{\changes{One-Bit} Precoding Design for Single-User Scenario}
\label{sec:prec_single}

This section proposes a practical algorithm for designing the
single-user \changes{one-bit} symbol-level precoder, or equivalently the
\changes{normalized} transmitting signal, $\bx_i$, corresponding to each symbol, $s^i$,
assuming fixed constellation range $c$. Using \eqref{SERu_final} as
the SER metric, it can be seen that the only term that depends on \changes{the precoder} for a fixed constellation is $\hat{d}_{ii} =\left|
\sqrt{\tfrac{P}{M}}\mathbf{h}^H\bx_i -s^i\right|$. Since the $Q$-function
is a decreasing function, the transmit signal design problem for
the $i^\text{th}$ symbol \changes{can be} written as
%29
\begin{equation}
\bx^{*}_i = \underset{\bx_i \in \mathcal{X}^M}{\operatorname{argmin}} ~~ \left| \sqrt{\tfrac{P}{M}}\mathbf{h}^H\bx_i -s^i\right|.
\label{precod_prob}
\end{equation}
Solving this optimization problem is hard due to the combinatorial nature of its constraints. Here, we seek to find a good solution for \eqref{precod_prob} with low complexity.
 
Toward this aim, we observe that for a fixed channel $\bh$, the possible
realizations of $\mathbf{h}^H\bx$, when $\bx \in \mathcal{X}^M$, are
densely distributed close to the origin; an example for $M=8$ is depicted in
Fig.~\ref{fig:scatter}. This observation suggests that if we can choose the
transmitting signal across a suitably chosen subset of antennas in a greedy fashion so
as to bring the residual to be small, then we can 
use exhaustive search  at the remaining (e.g. 8) antennas  to drive
the residual very close to zero. Based on this, we propose the following two-step algorithm to find a reasonable solution for \eqref{precod_prob}:

\begin{itemize}
\item \textbf{Step 1:} Use an iterative greedy approach to bring the
residual close to zero by designing the precoded signal at a suitably chosen subset of
$M_1 = M - M_2$ antennas where $M_1 \gg M_2$. In particular, at each
iteration, the algorithm selects one antenna and 
its corresponding transmitting signal 
to minimize the norm of the residual.
This procedure is executed until the
transmitting signals for $M_1$ antennas are all determined.  
\item \textbf{Step 2:} Design the transmitting signals of the other $M_2$
antennas to further approximate the desired signal by performing an exhaustive
search. Note that the exhaustive search is feasible as $M_2 \ll M$.
\end{itemize}
The mathematical details of the above two-step algorithm are illustrated in Algorithm~\ref{Alg:single_user}. The overall computational complexity of the algorithm is $O(MM_1)+ O(4^{M_2})$, where the first and the second terms correspond to the first and the second steps of the proposed algorithm, respectively. In our numerical results, we observe that $5 \le M_2 \le 10$ is a reasonable range of choices for $M_2$ which provides good performance with reasonable computational complexity.

 %%%%%%%%%%%%%%%%%%%%%%%%%%%%%%%%%%%
%Algorithm 1
%%%%%%%%%%%%%%%%%%%%%%%%%%%%%%%%%%%
\begin{algorithm}[t]
\caption{Single-User Transmitting Signal Design}
\label{Alg:single_user}
\textbf{Inputs:} $P$, $M$, $\bh$, $s^i$
\begin{algorithmic}
\State \textbf{\underline{Step 1:}}
\State Initialize the residual as $s_r = s^i$ and define the set $\mathcal{J} = \{ 1,\ldots,M\}$.
				   	\For{$j= 1 \to M_1$}
				   	\State $\left(j^*,x^*\right) = \underset{j\in \mathcal{J}, x \in \mathcal{X}}{\operatorname{argmin}} \left| s_r - \sqrt{\tfrac{P}{M}} {h}_j^{\dagger}x\right|$,
				   	\State $x_{j^*} = x^*$,
				   	\State $\mathcal{J} = \mathcal{J}\setminus j^{*}$,
				   	\State $s_r = s_r - \sqrt{\tfrac{P}{M}} {h}_{j^{*}}^{\dagger}x_{j^*}$,
          \EndFor
\State \textbf{\underline{Step 2:}}
\State 
Use the exhaustive search method to solve problem $\displaystyle\min_{x_j\in \mathcal{X}, \forall j\in \mathcal{J}} \Big| s_r - \sqrt{\tfrac{P}{M}} \sum_{j\in \mathcal{J}} h_j^\dagger x_j\Big |$.
\end{algorithmic}
\end{algorithm}
 
%%%%%%%%%%%%%%%%%%%%%%%%%%%%%%%%%%%
% VI) Constellation Range and One-Bit} Precoding Design for Multi-User Scenario
%%%%%%%%%%%%%%%%%%%%%%%%%%%%%%%%%%%
 \section{Constellation Range and \changes{One-Bit} Precoding Design for Multi-User Scenario}
 \label{sec:multi}
 This section generalizes the proposed designs for the QAM
constellation range and the transmitting signals in
Section~\ref{sec:const_single} and Section~\ref{sec:prec_single} from
the single-user scenario to the multi-user scenario.
 
 %%%%%%%%%%%%%%%%%%%%%%%%%%%%%%%%%%%
% VI.A) Constellation Range Design
%%%%%%%%%%%%%%%%%%%%%%%%%%%%%%%%%%%
 \subsection{Constellation Range Design}
\label{sec:range_multi}
When a multi-antenna BS serves multiple users at the same time, the constellation range depends not only on the channel, but also
on the number of users $K$ being served simultaneously and the
constellation size. This section provides an analysis of the optimal
constellation range in the massive MIMO regime under 
multi-user symbol-level  precoding. We begin by considering the infinite-resolution ZF scheme with per-symbol total power constraint across the antennas. 

Fixing a channel realization $\bH$, the ZF precoding vector can be found
by solving the following optimization problem which illustrates the
minimum required power so that the noiseless received signal of each
user is exactly equal to its intended signal, namely,
%30
\begin{subequations}
\label{c_inf_t_multi_prob}
\begin{eqnarray}
&\displaystyle{\min_{\bx}} &  \| \bx \|_2^2\\
&\text{s.t.} & \sqrt{\tfrac{P}{M}} \bH \bx = \bs.
\end{eqnarray}
\end{subequations}
This problem is convex and its optimal solution is given by 
%31
\begin{equation}
\bx^* = \sqrt{\tfrac{M}{P}}\bH^H(\bH\bH^H)^{-1} \bs.
\end{equation}

The constellation range should be designed such that for a vector of
intended constellation points, $\bs = [s_1,\ldots,s_K]$, the
\changes{normalized} transmitted signal satisfies $\| \bx^{*} \|_2^2 \leq M$. This is
equivalent to 
  %32
 \begin{equation}
  \bs^H (\bH\bH^H)^{-1} \bs \leq P.
  \label{cond_ZFIPC}
  \end{equation}

Now using the pseudo-orthogonal property of the channels of different
users in the large-scale massive MIMO regime, we can approximate
$(\bH\bH^H)^{-1} $ with a diagonal matrix in which the $k^\text{th}$
diagonal element is $\tfrac{1}{\| \bh_k\|_2^2}$. 
By approximating $(\bH\bH^H)^{-1}$ with that diagonal matrix, the inequality \eqref{cond_ZFIPC} can be rewritten as  
  %33
 \begin{equation}
   \sum_{k=1}^K \frac{|s_k|^2}{\| \bh_k \|_2^2} \leq P.
  \label{cond_ZFIPC_2}
  \end{equation}

 This constraint can be further simplified by taking advantage of the
channel hardening phenomenon in the massive MIMO regime that allows
the approximation $\|\bh_k\|_2^2 \approx M$ for all $k$ as
%34
 \begin{equation}
   \sum_{k=1}^K |s_k|^2 \leq PM.
  \label{cond_ZFIPC_3}
  \end{equation}
The constellation range design problem is then to choose the range $c$
so that the above constraint is satisfied for most 
realizations of \changes{$(s_1,s_2,\ldots,s_K)$} if they are 
within the
range.
   
Consider a particular user $k$ whose intended signal $s_k$ is selected in an
independent and identically distributed (i.i.d.) fashion from an $N^2$-QAM 
constellation with range $c$. A routine calculation shows that
 %35
\begin{equation}
\mu_s = \mathbb{E}\left\{ |s_k|^2 \right\} = \frac{N+1}{6(N-1)}c^2,
\label{eq:mu}
\end{equation}
and 
%36
\begin{equation}
\sigma^2_s = \mathbb{E}\left\{ \left( |s_k|^2 - \mu_s\right)^2 \right\} =  \frac{(N+1)(N^2-4)}{90(N-1)^3}c^4.
\label{eq:sigma}
\end{equation}
When a large number of users are precoded together, by the central
limit theorem, we have that
%37
\begin{equation}
\label{Kinfty_approx}
\sqrt{K}\left( \tfrac{1}{K}\sum_{k=1}^K |s_k|^2 - \mu_s \right) \overset{} \to \mathcal{N}\left(0,\sigma_s^2 \right).
\end{equation}

The optimal constellation range for ZF precoding is the maximum
range $c$ such that the precoded signal almost always
satisfies the per-symbol power constraint, i.e., \eqref{cond_ZFIPC_3}.
In this paper, we propose to design $c$ so that the mean of
$\sum_{k=1}^K |s_k|^2$ in \eqref{cond_ZFIPC_3} is within two
standard deviation from the constraint $PM$. This ensures that 
\eqref{cond_ZFIPC_3} is satisfied with \changett{high probability}.
Using \eqref{Kinfty_approx} and the properties of the Gaussian
distribution, this design goal of $K\mu_s + \changett{2}\sqrt{K}\sigma_s = PM$
yields that
%38
\begin{equation}
c^{*}_\text{ZF} = \sqrt{\frac{2PM}{f(K,N)}},
\label{c_inf_t_multi}
\end{equation}
where 
%39
\begin{equation}
f(K,N)= K \frac{N+1}{3(N-1)} +\changett{2}\sqrt{K\frac{(N+1)(N^2-4)}{22.5(N-1)^3}}.
\label{fKN}
\end{equation}
Comparing the constellation range design of the infinite-resolution ZF
for the multi-user scenario with that of the single-user case considered
in Section~\ref{sec:const_single}, i.e., \eqref{c_inf_t_multi} vs \eqref{c_inf_t_M}, 
it can be seen that there is an extra scaling factor $\sqrt{\tfrac{1}{f(K,N)}}$
in the constellation range expression when a massive MIMO BS
simultaneously serves large number of users.

Finally, we consider the constellation range design for the \changes{one-bit} precoding case.  Inspired by the form of
\eqref{c_inf_t_multi}, this paper proposes to multiply the same
scaling factor $\sqrt{\tfrac{1}{f(K,N)}}$ to our proposed single-user
constellation range design for \changes{one-bit} precoding in \eqref{c_1bit_single} to
account for the effect of serving multiple users, i.e., 
%40
\begin{eqnarray} 
\label{c_1bit_multi}
c^{*}_\text{$1$-bit} = \changett{\sqrt{2/\pi}}\sqrt{\frac{2PM}{f(K,N)}},
\end{eqnarray}
Numerical results are presented later in the paper to show that this is an appropriate design. 

%%%%%%%%%%%%%%%%%%%%%%%%%%%%%%%%%%%
% VI.B) One-Bit Precoding Design
%%%%%%%%%%%%%%%%%%%%%%%%%%%%%%%%%%%
 \subsection{\changes{One-Bit} Precoding Design} 

We now present a practical algorithm for \changes{one-bit symbol-level precoder
design} for the multi-user MISO system. 
In the multi-user scenario where the intended symbol of each user is selected from a \changes{square} $N^2$-QAM constellation, the vector of intended symbols has $N^{2K}$ choices. For each choice of $\bs^i$ where $ i\in\{1,\ldots,N^{2K}\}$, we seek to design the \changes{normalized} transmitting signal $\bx_i$ to minimize the average user SER, mathematically,
 %41
 \begin{equation}
 \label{prob_1bit_multi_orig}
 \min_{\bx_i \in \mathcal{X}^M} \sum_k \frac{g_k^i}{N^2} Q\left( \frac{d_\text{$1$-bit} - 2\left|\sqrt{\tfrac{P}{M}}\mathbf{h}_k^H\bx_i - s_k^i\right|} {2\sigma} \right),
 \end{equation}
 where $d_\text{$1$-bit} = \tfrac{c^{*}_\text{$1$-bit}}{N-1}$ and $s_k^i$ is the $k^\text{th}$ element of $\bs^i$. 
Instead of tackling the problem in \eqref{prob_1bit_multi_orig}, due
to the rapidly decreasing shape of the $Q$-function, we can consider
the following optimization problem 
 %42
  \begin{equation}
 \label{prob_1bit_multi}
 \min_{\bx_i \in \mathcal{X}^M}
\left\|\sqrt{\tfrac{P}{M}}\mathbf{H}\bx_i - \bs^i\right\|_\infty,
 \end{equation}
which minimizes the maximum approximation error across all the users. 
For solving this problem, we use a generalization of the two-step 
algorithm proposed in Section~\ref{sec:prec_single} for the
single-user case. 

The proposed two-step approach is summarized in Algorithm~\ref{Alg:multi_user}.
Specifically, the first step of the proposed algorithm seeks to bring the residual vector close to the origin, i.e., minimize the $2$-norm of the residual symbol vector, by choosing the transmit values over a subset of $M_1$ 
antennas in a greedy fashion. In the second step of the algorithm, exhaustive search is performed over
the remaining $M_2 = M-M_1$ antennas to minimize the maximum deviation from the residual symbols, i.e., to minimize the $\infty$-norm of the residual symbol vector. The rational behind minimizing the $2$-norm in the first step is that the realizations of $\sqrt{\tfrac{P}{M}}\mathbf{H}\bx$ for $\bx \in \mathcal{X}^M$ are distributed densely close to the origin which means that the vectors with smaller $2$-norm are surrounded by more realizations. Therefore, it is expected that the second step of the algorithm achieves a better performance if the residual symbol vector that is passed to it as an input has a smaller $2$-norm.

%%%%%%%%%%%%%%%%%%%%%%%%%%%%%%%%%%%
%Algorithm 2
%%%%%%%%%%%%%%%%%%%%%%%%%%%%%%%%%%%
\begin{algorithm}[t]
\caption{Multi-User Transmitting Signal Design}
\label{Alg:multi_user}
\textbf{Inputs:} $P$, $M$, $\bH$, $\bs^i$
\begin{algorithmic}
\State \textbf{\underline{Step 1:}}
\State Initialize the residual symbol vector as $\bs_r = \bs^i$ , define the set $\mathcal{J} = \{ 1,\ldots,M\}$, and denote the $j^\text{th}$ column of $\bH$ by $\tilde{\bh}_j$.
				   	\For{$j= 1 \to M_1$}
				   	\State $\left(j^*,x^*\right) = \underset{j\in \mathcal{J}, x \in \mathcal{X}}{\operatorname{argmin}} \left\| \bs_r - \sqrt{\tfrac{P}{M}} \tilde{\bh}_j x \right\|_2$,
				   	\State $x_{j^*} = x^*$,
				   	\State $\mathcal{J} = \mathcal{J}\setminus j^{*}$,
				   	\State $\bs_r = \bs_r - \sqrt{\tfrac{P}{M}} \tilde{\bh}_{j^{*}} x_{j^*}$,
          \EndFor
\State \textbf{\underline{Step 2:}}
\State 
 Use the exhaustive search method to solve problem $\displaystyle\min_{x_j\in \mathcal{X}, \forall j\in \mathcal{J}} \Big\| \bs_r - \sqrt{\tfrac{P}{M}} \sum_{j\in \mathcal{J}}\tilde{\bh}_j x_j\Big \|_\infty$.
\end{algorithmic}
\end{algorithm}  

The computational complexity of the first and the second steps of
Algorithm~\ref{Alg:multi_user} are $O(KMM_1)$ and $O(K4^{M_2})$,
respectively.  Simulation results presented in Section~\ref{sec:sim}
show excellent numerical performance of the algorithm with reasonable
complexity.
 
%%%%%%%%%%%%%%%%%%%%%%%%%%%%%%%%%%%
% VII) Performance Gap of One-Bit Precoding vs. Conventional ZF
%%%%%%%%%%%%%%%%%%%%%%%%%%%%%%%%%%%
\subsection{Performance Gap of \changes{One-Bit} Precoding vs. \changes{Infinite-Resolution ZF}}
\label{sec:anal_compare}

For infinite-resolution ZF \changes{under the instantaneous power constraint} with the constellation range designed as \eqref{c_inf_t_multi}, since almost surely the constellation points
can be reconstructed exactly, the symbol error rate simply scales with
the minimum distance as follows:
%43
\begin{equation}
\operatorname{SER} = \bar{g}_N Q\left( \frac{d}{2\sigma}\right),
\label{SER_general}
\end{equation}  
where $\bar{g}_N = 4\left(1 - \tfrac{1}{N}\right)$ is the average number of nearest \changes{neighbors} in the symbol constellation, and $d$ is the minimum distance in the constellation,
which can be expressed as:
%44
\begin{equation}
 d^*_\text{ZF} = \frac{c^{*}_\text{ZF}}{N-1}= \sqrt{\frac{2PM}{\tilde{f}(K,N)}},
\end{equation}
where $\tilde{f}(K,N) = (N-1)^2 f(K,N)$.

 In the \changes{one-bit} precoding scheme, there is no guarantee that
the noiseless received signals can exactly realize the intended
symbols. However, if we design the constellation range carefully so
that the \changes{one-bit} precoder can approximate the noiseless receive
signals to be very close to the intended signals, then the term
$2\hat{d}_{ii}$ in \eqref{SERu_final} is negligible as compared to the
minimum distance $d$, and we can tightly approximate the SER of the
proposed scheme using the same equation as \eqref{SER_general} with 
 %45
 \begin{equation}
 d_\text{$1$-bit} = \frac{c^{*}_\text{$1$-bit}}{N-1}= \changett{\sqrt{2/\pi}}\sqrt{\frac{2PM}{\tilde{f}(K,N)}}.
 \end{equation}

This suggests that
the \changes{one-bit} precoding scheme requires about $10\log_{10}{\tfrac{1}{\changett{{2/\pi}}}} \approx 2$dB more power than the
conventional ZF to achieve the same performance \changes{in the massive MIMO regime}. The simulation
results of this paper indeed show this $2$dB gap under the proposed
constellation range design, thereby verifying that the proposed design
is a reasonable one. 
 
We can also characterize the performance gap of \changes{one-bit} precoding and
conventional ZF in terms of the number of extra antennas that would
need to be added to \changes{one-bit} precoding in order to achieve the same
performance as conventional ZF. Using \eqref{SER_general}, it can be
seen that
%46
\begin{equation}
  M_\text{$1$-bit} = \tfrac{1}{\changett{{2/\pi}}} M_\text{ZF} \approx 1.57 M_\text{ZF}. 
\end{equation}
  This means that \changes{one-bit} precoding with about $50\%$ more number of
antennas can achieve the same performance as infinite-resolution ZF
precoding under the same instantaneous per-symbol power constraint. 
 
It is worth mentioning that although the above discussion pertains to
the multi-user case, since the same constellation range design, i.e.,
$c^{*}_\text{$1$-bit}= \changett{\sqrt{2/\pi}} c^{*}_\text{inf,t}$ is used for the
single-user case, the performance gap between the infinite-resolution
precoding with per-symbol total power constraint and the proposed
\changes{one-bit} precoding design for the single-user case is also about $2$dB,
which again translates to requiring about $50\%$ more
antennas for \changes{one-bit} precoding to achieve the same performance as 
infinite-resolution precoding.
 
%%%%%%%%%%%%%%%?%%%%%%%%%%%%%%%%%%%%%
% VII) Simulations
%%%%%%%%%%%%%%%%%%%%%%%%%%%%%%%%%%%%
\section{Simulation Results}
\label{sec:sim}
In this section, we present numerical simulation results to support
the proposed design methodology for constellation range and to
evaluate the performance of the proposed algorithm for \changes{one-bit}
symbol-level precoding for both single-user and multi-user scenarios.

%%%%%%%%%%%%%%%%%%%%%%%%%%%%%%%%%%%
%Figure 5
%%%%%%%%%%%%%%%%%%%%%%%%%%%%%%%%%%%
\begin{figure}[t]
	\centering
	{\includegraphics[width=0.5\textwidth]{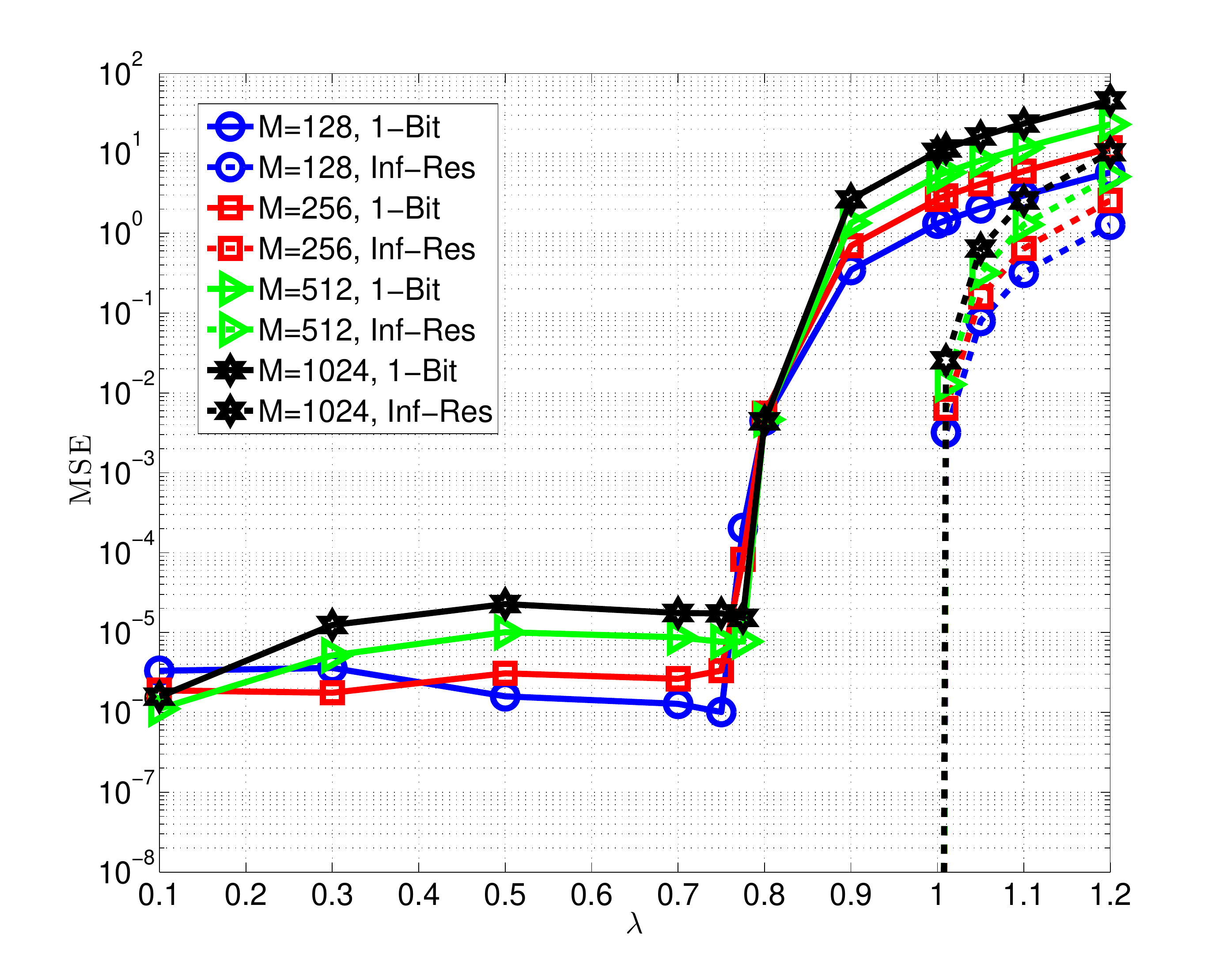}}
	\caption{Average MSE versus $\lambda$ in a single-user system for different numbers of antennas, $M$, when $P=1$ and $M_2 = 8$.}
	\label{fig:mse1}
\end{figure}

%%%%%%%%%%%%%%%?%%%%%%%%%%%%%%%%%%%%%
%  VII-A) MSE Analysis for Constellation Range Design
%%%%%%%%%%%%%%%%%%%%%%%%%%%%%%%%%%%%
 \subsection{MSE Analysis for Constellation Range Design}

First, we show that the constellation range design proposed in
Section~\ref{sec:1bit_range_single} for the single-user case, 
which sets $c^{*}_{\text{$1$-bit}} = \changett{\sqrt{2/\pi}}\sqrt{2P}\| \bh\|_2$  is
nearly optimal. Consider a BS with $M$ antennas serving one user by
transmitting symbols from a $16$-QAM constellation with constellation
range of $c$.  We evaluate the average MSE defined in (\ref{MSE_exp})
against the parameter
%47
\begin{equation}
\lambda = \frac{c}{ \sqrt{2P}\|\bh\|_2},
\label{eq:lambda}
\end{equation}
where the average is over the channel realizations and \changes{one-bit}
precoding is performed using Algorithm~\ref{Alg:single_user}.
The transmit power is set to $P =1$. The number of antennas $M$
ranges from 128 to 1024. Note that $M_2$ is set to $8$ in all cases in
order to keep the complexity manageable. 

Fig.~\ref{fig:mse1} shows that there is a sharp phase transition in
MSE for both infinite-resolution and \changes{one-bit} precoding. The phase
transition occurs at $\lambda=1$ for infinite-resolution precoding
with instantaneous per-symbol total power constraint, thus verifying 
the constellation range design (\ref{c_inf_t}), i.e.,
$c^{*}_{\text{inf,t}} = \sqrt{2P}\| \bh\|_2$.  
For \changes{one-bit} precoding, the phase transition occurs at \changett{about $\lambda= {\sqrt{2/\pi}} \approx 0.8$}, thus
verifying the constellation range design (\ref{c_1bit_single}), i.e., 
$c^{*}_{\text{$1$-bit}} = \changett{\sqrt{2/\pi}}\sqrt{2P}\| \bh\|_2$ even with the
proposed low-complexity practical \changes{one-bit} precoding algorithm. 

%%%%%%%%%%%%%%%%%%%%%%%%%%%%%%%%%%%
%Figure 6
%%%%%%%%%%%%%%%%%%%%%%%%%%%%%%%%%%%
\begin{figure}[t]
	\centering
	{\includegraphics[width=0.5\textwidth]{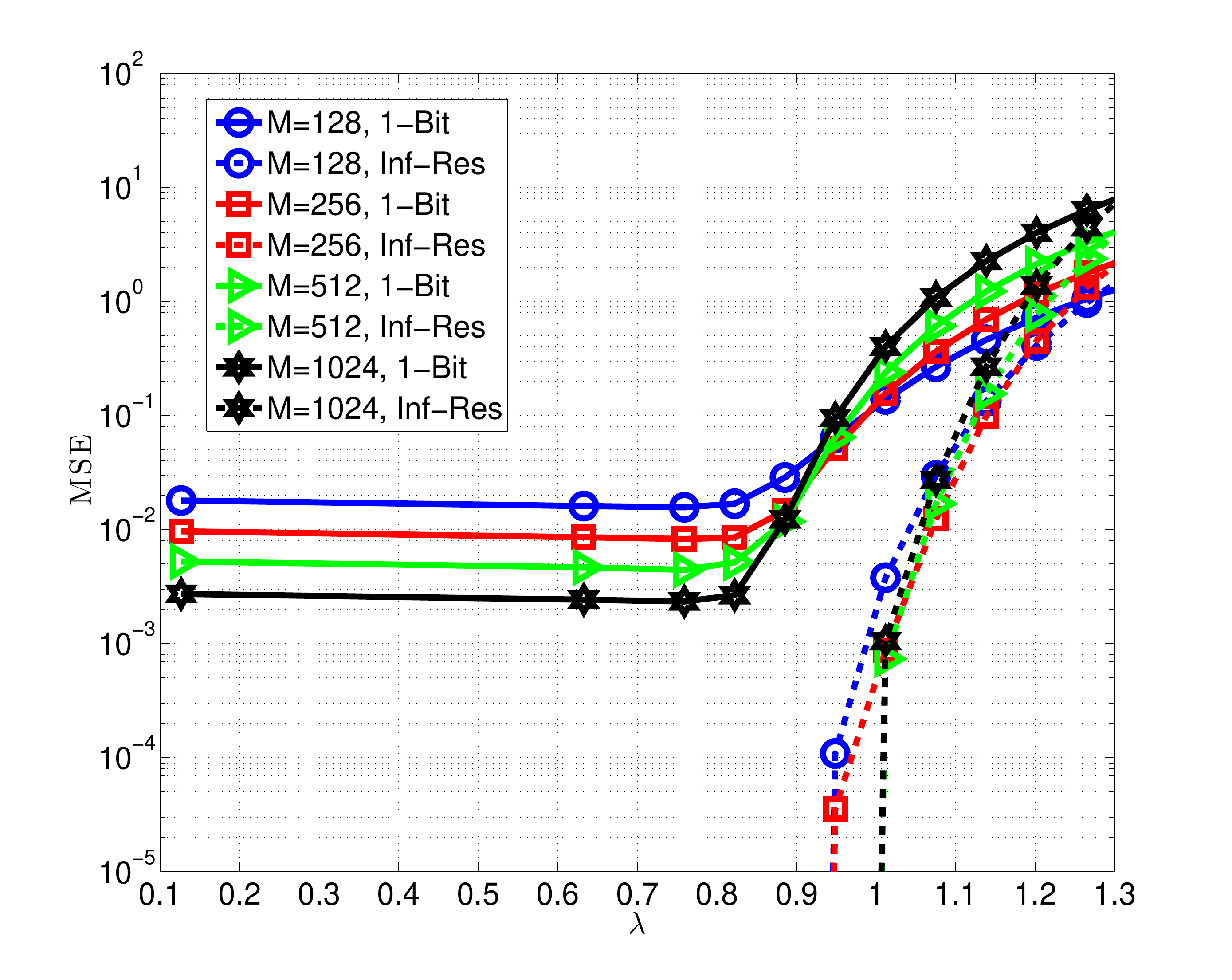}}
	\caption{Average MSE versus $\lambda$ in a $8$-user system for different numbers of antennas, $M$, when $P=1$ and $M_2 = 8$.}
	\label{fig:mse2}
\end{figure}

\changet{In Fig.~\ref{fig:mse1}, the MSE for the values of $\lambda$ smaller
than the phase transition point indicates the quality of the \changes{one-bit}
precoder design. One may ask whether the MSE achieved by the proposed
\changes{one-bit} precoding is sufficiently small so that it does not
significantly degrade the SER performance.  The answer is affirmative under typical system parameter ranges. 
Fig.~\ref{fig:mse1} suggests that for the transmit power budget $P=1$
the MSE about $10^{-5}$ can be achieved by the proposed algorithm.
Since the MSE scales with the power budget, the MSE of the proposed
algorithm with the power budget $P$ is about $10^{-5}P$.
Using the SER expression in \eqref{SERu_final} and observing that the
MSE is nearly uniform for all symbols before the phase transition, it
can be shown that 
%48
\begin{equation}
\operatorname{SER} \approx \bar{g}_N Q\left(
\frac{c^{*}_{\text{$1$-bit}}}{2\sigma(N-1)} - \frac{\sqrt{\operatorname{MSE}}}{\sigma} \right).
\end{equation}
  
  For the typical operating regime of SER (e.g. $10^{-3}-10^{-6}$), the contribution of the second term in the $Q$-function can be ignored if it is at least one order of magnitude smaller than the first term, i.e., 
   %49
  \begin{equation}
  0.1\frac{c^{*}_{\text{$1$-bit}}}{2\sigma(N-1)} \leq \frac{\sqrt{\operatorname{MSE}}}{\sigma}.
 \end{equation}
     For  $\operatorname{MSE} = 10^{-5}P$
and $c^{*}_{\text{$1$-bit}}= \changet{\sqrt{2/\pi}} \sqrt{2P}\| \bh\|_2 \approx 0.8
\sqrt{2PM}$, this condition translates to an upper-bound on the number of constellation points as: 
%50
\begin{equation}
N \leq 17.9 \sqrt{M} +1.
\label{N_cal_single}
\end{equation}

This condition is clearly satisfied in a typical single-user MISO system, with for example 128 antennas and with constellation size at most $N^2 = 2^{12}$. Therefore, the MSE of the proposed \changes{one-bit} precoding has negligible influence on the SER in such a system.}

%%%%%%%%%%%%%%%%%%%%%%%%%%%%%%%%%%%
%Figure 7
%%%%%%%%%%%%%%%%%%%%%%%%%%%%%%%%%%%
\begin{figure}[t]
	\centering
	{\includegraphics[width=0.5\textwidth]{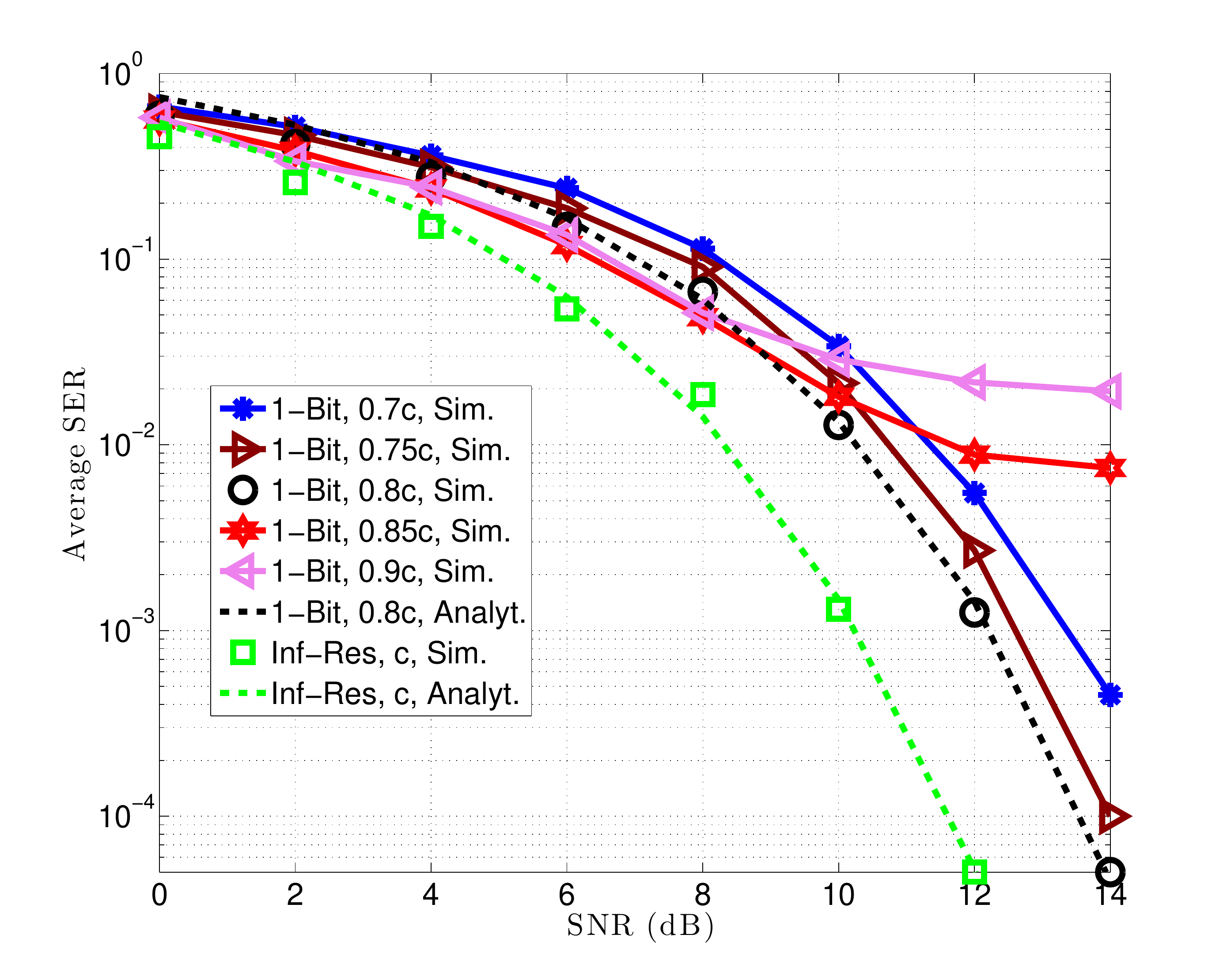}}
	\caption{Average SER versus SNR for different methods in a single-user MISO system with $M=256$ and $N=16$.}
	\label{fig:sim_single}
\end{figure}

Next, we show that the constellation range design for the multi-user
case proposed in Section \ref{sec:range_multi}, which sets
$c^{*}_\text{$1$-bit} = \changett{\sqrt{2/\pi}} \sqrt{\tfrac{2PM}{f(K,N)}} $ is also
nearly optimal. Toward this aim, we consider a $8$-user system with
transmit power $P=1$ and $16$-QAM signaling, i.e., $N=4$. We then vary
the constellation range of $16$-QAM constellation and plot the average
MSE, which for the multi-user setup is defined as 
%51
\begin{equation}
 \operatorname{MSE}(\mathbf{s}) = \mathbb{E}_\bH\left\{\left| \sqrt{\tfrac{P}{M}}\bH\bx - \bs\right|^2\right\},
 \label{MSE_exp_multi}
 \end{equation}
 against the parameter $\lambda$ defined as
 %52
\begin{equation}
\lambda = \frac{c}{ \sqrt{\frac{2PM}{f(K,N)}}}.
\label{eq:lambda}
\end{equation}

Fig.~\ref{fig:mse2} shows that for the multi-user scenario, there is
a phase transition at $\lambda$ very close to $1$ for
infinite-resolution ZF precoding, and $\changett{\sqrt{2/\pi} \approx 0.8}$ for \changes{one-bit} precoding, 
respectively. This verifies our
design choices (\ref{c_inf_t_multi}) and (\ref{c_1bit_multi}), i.e., $c^{*}_\text{ZF} = \sqrt{\tfrac{2PM}{f(K,N)}}$ and
$c^{*}_\text{$1$-bit} = \changett{\sqrt{2/\pi}} \sqrt{\tfrac{2PM}{f(K,N)}}$.

%%%%%%%%%%%%%%%?%%%%%%%%%%%%%%%%%%%%%
%  VII-A) SER Analysis for One-Bit Precoding
%%%%%%%%%%%%%%%%%%%%%%%%%%%%%%%%%%%%
\subsection{\changes{SER Analysis for One-Bit Precoding Design}}
We now evaluate the SER performance of the proposed one-bit precoding
scheme.  In the following simulations, 
the performance of different methods are
evaluated using the empirical average SER of the users calculated by averaging the SER over $10^3$ channel realizations and in
each realization $200$ symbols are transmitted, i.e., the coherence time of the channel is 
$200$ symbol transmissions. Further, the signal-to-noise ratio is defined as $\operatorname{SNR} = 10\log_{10} \left( {\tfrac{P}{2\sigma^2}} \right)$.

First, we evaluate a single-user communication setup with $256$-QAM,
i.e., $N=16$. We set the parameter $M_2=8$ in the proposed \changes{one-bit}
precoding algorithm.  The BS is assumed to be equipped with $M=256$ antennas. 
In Fig.~\ref{fig:sim_single}, the performance of the proposed \changes{one-bit}
precoding method with different constellation range designs is
compared to the infinite-resolution precoding benchmark with
per-symbol power constraint and constellation range of $c = \sqrt{2P}\| \bh\|_2$. 
It can be seen from Fig.~\ref{fig:sim_single} that 
the constellation range of $\changett{\sqrt{2/\pi}c \approx 0.8c}$ achieves the overall best performance, 
justifying our proposed design (\ref{c_1bit_single}) for \changes{one-bit} precoding. 

Fig.~\ref{fig:sim_single} \changes{also} shows that for the \changes{one-bit} precoding methods
with constellation range of larger than $0.8c$, there is an error floor 
in the large SNR regime. This is because in the large SNR regime, the 
average SER is dominated by the worst-case symbols, which are the 
corner constellation points. One-bit precoding with constellation
range larger than $0.8c$ leads to high reconstruction MSE for the corner points. 

%%%%%%%%%%%%%%%%%%%%%%%%%%%%%%%%%%%
%Figure 8
%%%%%%%%%%%%%%%%%%%%%%%%%%%%%%%%%%%
\begin{figure}[t]
	\centering
	{\includegraphics[width=0.5\textwidth]{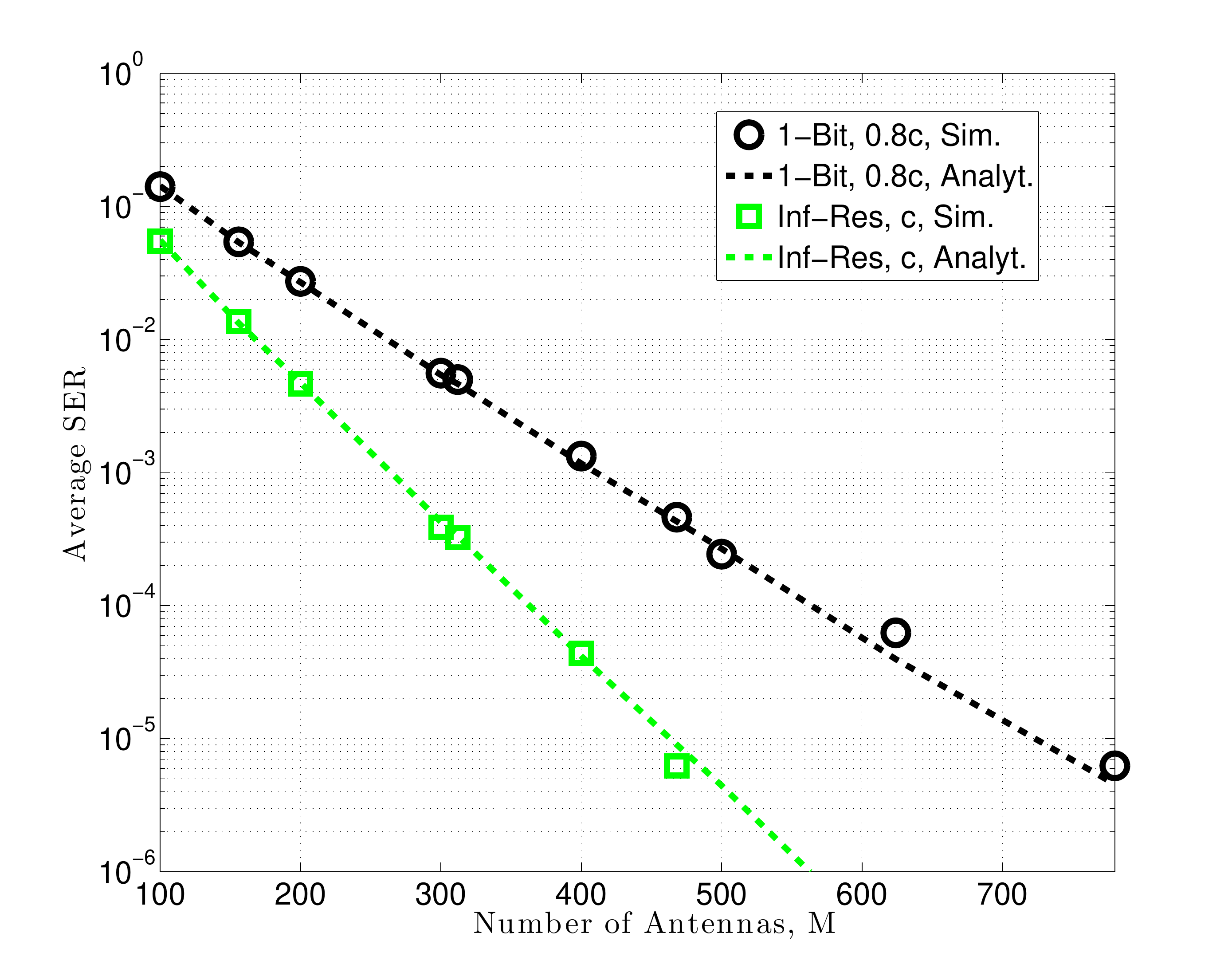}}
	\caption{Average SER versus the number of antennas, $M$, for different methods in a single-user MISO system with $\operatorname{SNR}=-4$dB and $N=4$.}
	\label{fig:sim_single_antennas}
\end{figure}

Moreover, Fig.~\ref{fig:sim_single} shows that the analytic SER expressions provided
in Section~\ref{sec:anal_compare} for the infinite-resolution precoding and the proposed \changes{one-bit} precoding with constellation range of $0.8c$ perfectly match the numerical simulation results and  the performance gap between  the proposed method and
the infinite-resolution case is about $2$dB, as predicted by the SER analysis. 

Next, we consider a single-user setup in which $16$-QAM constellation is employed and the SNR is set to be $-4$dB. For such a system, Fig.~\ref{fig:sim_single_antennas} plots the average SER  against the number of antennas at the BS for the proposed \changes{one-bit} precoding method as well as the infinite-resolution precoding method. It can be observed from Fig.~\ref{fig:sim_single_antennas} that the proposed \changes{one-bit} precoding design can achieve the same performance as the infinite-resolution precoding if the BS in the \changes{one-bit} precoding architecture is equipped with about $50\%$ more number of antennas as compared to the infinite-resolution case, as predicated by the SER analysis in Section~\ref{sec:anal_compare}.

\begin{figure}[t]
	\centering
	{\includegraphics[width=0.5\textwidth]{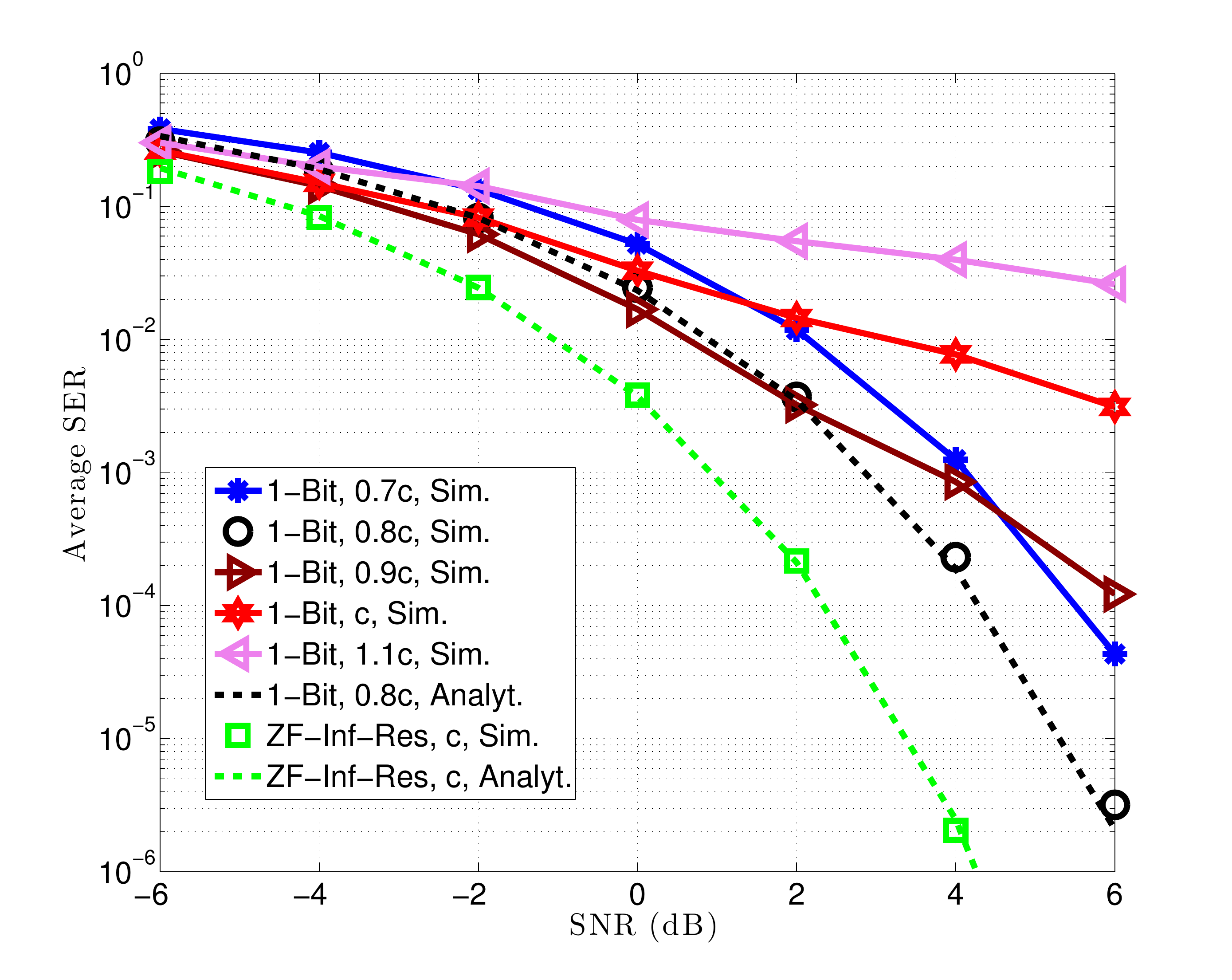}}
	\caption{Average SER versus SNR for different methods in a $8$-user MISO system with $M=512$ and $N=4$.}
	\label{fig:sim_multi}
\end{figure}

Finally, we consider a multi-user scenario in which $K=8$ users are
served with a BS equipped with $512$ antennas using 16-QAM, i.e.,
$N=4$. The performance of the proposed one-bit precoding method with
$M_2=8$ and different design choices for constellation range is
evaluated as compared to the infinite-resolution ZF benchmark with
constellation range of $c = \sqrt{\tfrac{2PM}{f(K,N)}}$ described in
Section~\ref{sec:multi}.
Fig.~\ref{fig:sim_multi} plots the average SER against the SNR. 

It can be observed from  Fig.~\ref{fig:sim_multi} that the analytic
SER expression presented in Section~\ref{sec:anal_compare} accurately
characterizes the average SER in the simulation for the infinite-resolution ZF precoding as well as the proposed \changes{one-bit} precoding with $\changett{\sqrt{2/\pi}c \approx 0.8c}$.
Further, similar to
the single-user case, a performance gap of $2$dB between the proposed
method with constellation range $0.8c$ and conventional ZF is observed in
Fig.~\ref{fig:sim_multi}. Finally, the \changes{one-bit} precoding
schemes with constellation range larger than the proposed design of
$0.8c$ 
all suffer from an error floor in the high SNR regime, leading to the
conclusion that $0.8c$ is
an appropriate design for the constellation range.

%%%%%%%%%%%%%%%%%%%%%%%%%%%%%%%%%%%
%Figure 9
%%%%%%%%%%%%%%%%%%%%%%%%%%%%%%%%%%%

%%%%%%%%%%%%%%%?%%%%%%%%%%%%%%%%%%%%%
%% Conclusion
%%%%%%%%%%%%%%%%%%%%%%%%%%%%%%%%%%%%
\section{Conclusion}
\label{sec:con}

This paper considers the \changes{one-bit} symbol-level precoding architecture
for a downlink massive MIMO system. First, we consider the problem of
designing the QAM constellation range and the precoder for the
single-user scenario in order to minimize the SER. We propose to set
the QAM constellation range of \changes{one-bit} precoding as the optimal
constellation range of infinite-resolution precoding reduced by the
factor of $\changett{\sqrt{2/\pi}}$ or about $0.8$.  We then propose a two-step heuristic algorithm to
design the precoder, which enjoys a low complexity and exhibits
excellent numerical performance. We also generalize the proposed
designs for the multi-user scenario. \changes{In particular, we first propose} constellation range
design for the infinite-resolution ZF case then further scale it by
$\changett{\sqrt{2/\pi}}$ for the \changes{one-bit} precoding case. Finally, we analytically study the
performance of the proposed scheme and show that for large-scale
antenna arrays, there is \changes{a} $2$dB gap between the proposed
design and the conventional ZF scheme with per-symbol power
constraint. The simulation results verify that the proposed design can
achieve a promising performance for large-scale antenna arrays with
low-resolution DACs.

%%%%%%%%%%%%%%%%%%%%%%%%%%%%%%%%%%%%
%% Refrences
%%%%%%%%%%%%%%%%%%%%%%%%%%%%%%%%%%%%
\bibliographystyle{IEEEtran}
\bibliography{IEEEabrv,refrences}

%%%%%%%%%%%%%%%%%%%%%%%%%%%%%%%%%%%%
%% Biography
%%%%%%%%%%%%%%%%%%%%%%%%%%%%%%%%%%%%
%%%%%%%%%%%%%%%%%% Foad Sohrabi
\begin{IEEEbiography}[{\includegraphics[width=1in,height=1.25in,clip,keepaspectratio]{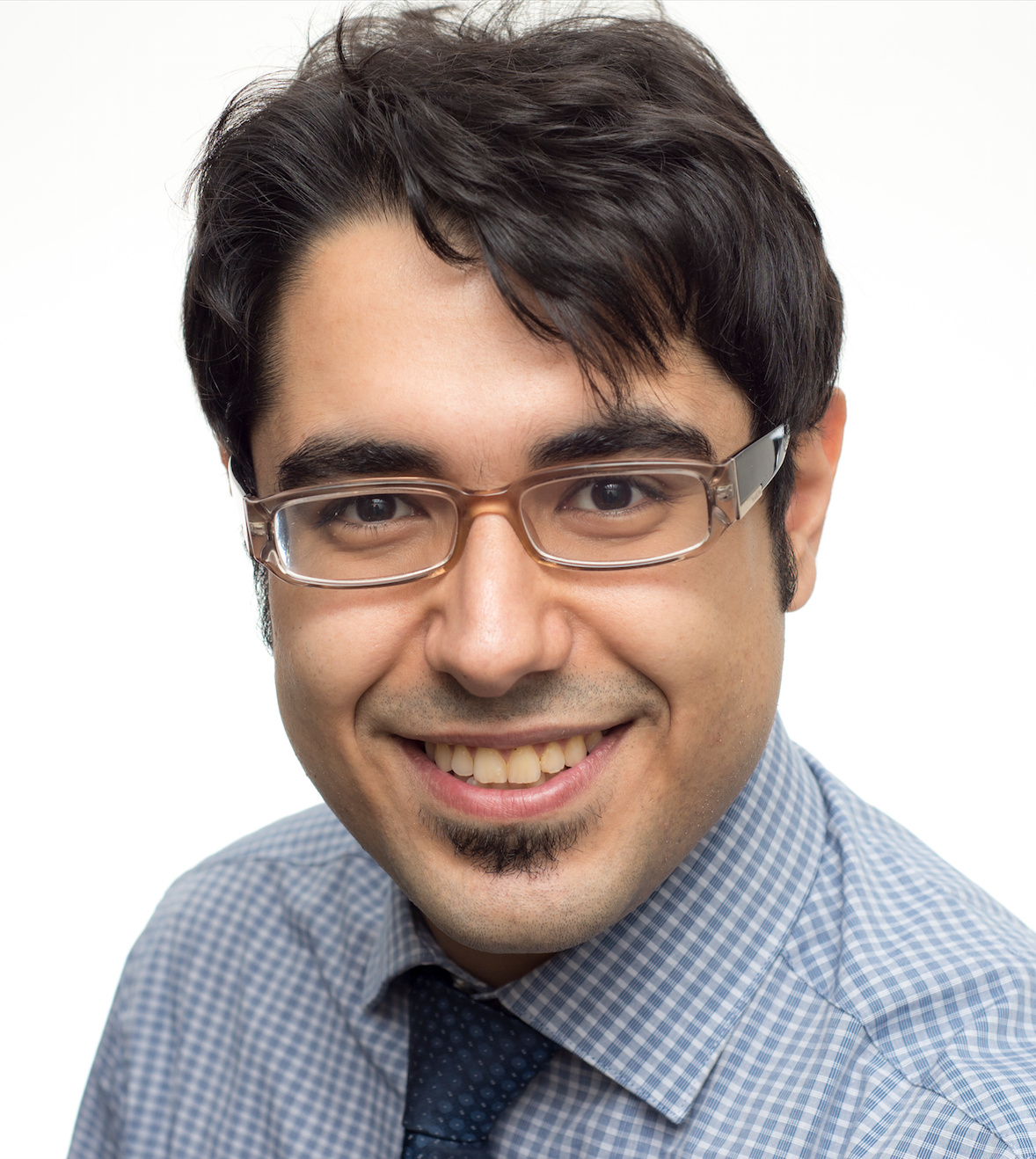}}]{Foad Sohrabi}
(S'13) received his B.A.Sc.\ degree in 2011 from University of Tehran, Tehran, Iran, his M.A.Sc.\ degree in 2013 from McMaster University, Hamilton, ON, Canada, and his Ph.D.\ degree in 2018 from University of Toronto, Toronto, ON, Canada, all in Electrical and Computer Engineering. During his PhD study, he was a research intern at Bell-Labs, Alcatel-Lucent, in Stuttgart, Germany for 6 months. 
He is currently a Postdoctoral Fellow with University
of Toronto, Toronto, ON, Canada.
His main research interests include MIMO communications, optimization theory, wireless communications, and signal processing. He received an IEEE Signal Processing Society Best Paper Award in 2017.
\end{IEEEbiography}

%%%%%%%%%%%%%%%%%% Ya-Feng Liu
\begin{IEEEbiography}[{\includegraphics[width=1in,height=1.25in,clip,keepaspectratio]{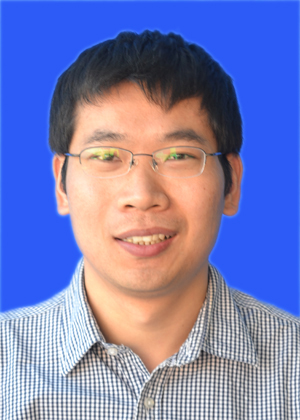}}]
{Ya-Feng Liu} (M'12) received the B.Sc. degree in applied mathematics in 2007 from Xidian University, Xi'an, China, and the Ph.D. degree in computational mathematics in 2012 from the Chinese Academy of Sciences (CAS), Beijing, China. During his Ph.D. study, he was supported by the Academy of Mathematics and Systems Science (AMSS), CAS, to visit Professor Zhi-Quan (Tom) Luo at the University of Minnesota (Twins Cities) from February 2011 to February 2012. After his graduation, he joined the Institute of Computational Mathematics and Scientific/Engineering Computing, AMSS, CAS, Beijing, China, in July 2012, where he is currently an Assistant Professor. His main research interests are nonlinear optimization and its applications to signal processing, wireless communications, and machine learning. He is especially interested in designing efficient algorithms for optimization problems arising from the above applications.

Dr. Liu has served as a guest editor of the Journal of Global Optimization. He is a recipient of the Best Paper Award from the IEEE International Conference on Communications (ICC) in 2011 and the Best Student Paper Award from the International Symposium on Modeling and Optimization in Mobile, Ad Hoc and Wireless Networks (WiOpt) in 2015.
\end{IEEEbiography}

%%%%%%%%%%%%%%%%%%% We YU
\begin{IEEEbiography}[{\includegraphics[width=1in,height=1.25in,clip,keepaspectratio]{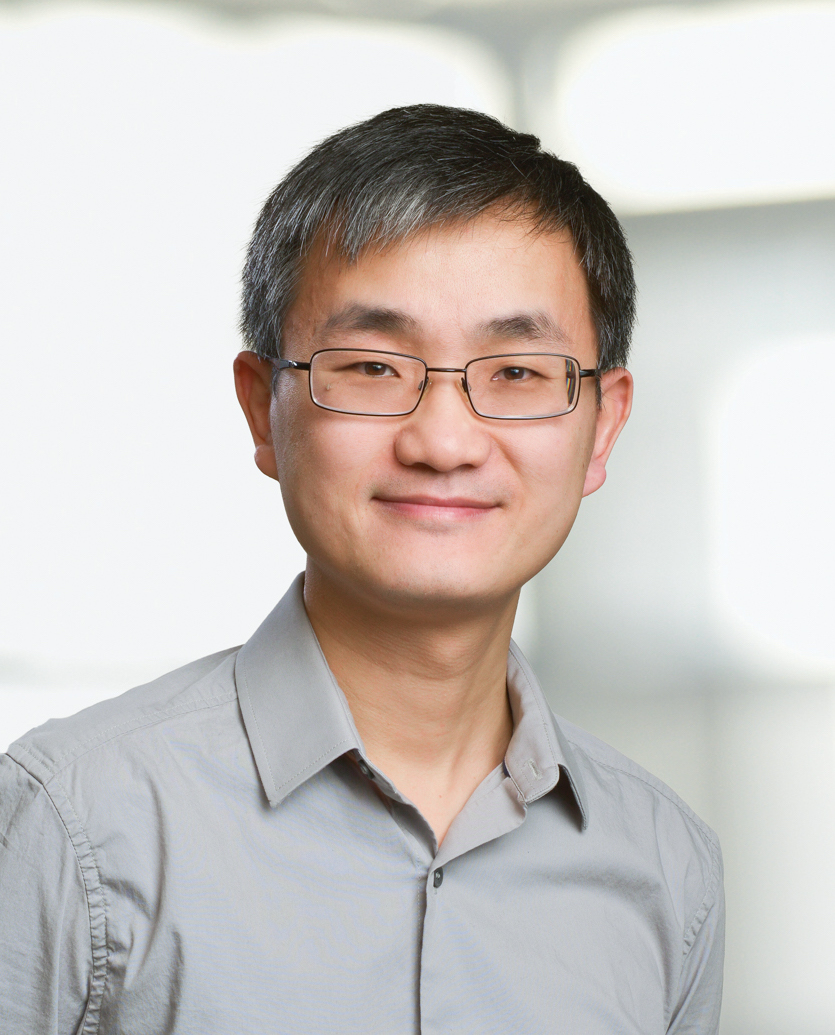}}]
{Wei Yu} (S'97-M'02-SM'08-F'14) received the B.A.Sc. degree in Computer Engineering and Mathematics from the University of Waterloo, Waterloo, Ontario, Canada in 1997 and M.S. and Ph.D. degrees in Electrical Engineering from Stanford University, Stanford, CA, in 1998 and 2002, respectively. Since 2002, he has been with the Electrical and Computer Engineering Department at the University of Toronto, Toronto, Ontario, Canada, where he is now Professor and holds a Canada Research Chair (Tier 1) in Information Theory and Wireless Communications. His main research interests include information theory, optimization, wireless communications and broadband access networks.

Prof. Wei Yu currently serves on the IEEE Information Theory Society Board of Governors (2015-20). He was an IEEE Communications Society Distinguished Lecturer (2015-16). He is currently an Area Editor for the IEEE Transactions on Wireless Communications (2017-20). He served as an Associate Editor for IEEE Transactions on Information Theory (2010-2013), as an Editor for IEEE Transactions on Communications (2009-2011), and as an Editor for IEEE Transactions on Wireless Communications (2004-2007). He is currently the Chair of the Signal Processing for Communications and Networking Technical Committee of the IEEE Signal Processing Society (2017-18) and served as a member in 2008-2013. Prof. Wei Yu received the Steacie Memorial Fellowship in 2015, the IEEE Signal Processing Society Best Paper Award in 2017 and 2008, a Journal of Communications and Networks Best Paper Award in 2017, an IEEE Communications Society Best Tutorial Paper Award in 2015, an IEEE ICC Best Paper Award in 2013, the McCharles Prize for Early Career Research Distinction in 2008, the Early Career Teaching Award from the Faculty of Applied Science and Engineering, University of Toronto in 2007, and an Early Researcher Award from Ontario in 2006. Prof. Wei Yu is a Fellow of the Canadian Academy of Engineering, and a member of the College of New Scholars, Artists and Scientists of the Royal Society of Canada. He is recognized as a Highly Cited Researcher.
\end{IEEEbiography}

\end{document}